\documentclass[reprint,aps,amsmath,amssymb,amsfonts,showkeys]{revtex4-2}
\usepackage[T1]{fontenc}
\usepackage[pdftex]{graphicx,xcolor}
\usepackage{dcolumn}
\usepackage{times}
\usepackage{multirow}
\usepackage{physics}
\usepackage{siunitx}
\usepackage{here}
\AtBeginDocument{\RenewCommandCopy\qty\SI}

\usepackage{xspace}
\usepackage{ulem}

\newcommand{\sro}{Sr$_2$RuO$_4$\xspace}
\newcommand{\sroru}{Sr$_2$RuO$_4$-Ru\xspace}

\usepackage[pdftex,bookmarks=false,bookmarksnumbered=false,colorlinks=true,linkcolor=black,citecolor=blue,filecolor=black,urlcolor=blue]{hyperref}

\begin{document}
\title{Anisotropic thermoelectric transport in the Sr$_{2}$RuO$_{4}$-Ru eutectic system}

\author{Takeru~Ito$^1$}
\email{6224504@alumni.tus.ac.jp}
\author{Shogo~Yoshida$^1$}
\email{s_yoshida@rs.tus.ac.jp}
\author{Hiroshi Yaguchi$^1$}
\author{Ryuji~Okazaki$^1$}
\email{okazaki@rs.tus.ac.jp}
\affiliation{$^1$Department of Physics and Astronomy, Tokyo University of Science, Noda 278-8510, Japan}

\begin{abstract}
The \sroru eutectic system, in which metallic Ru inclusions are embedded in a \sro matrix, exhibits an enhanced superconducting onset temperature of about 3~K, known as the 3-K phase.
Here we investigate the in-plane thermoelectric response of this eutectic system in the normal state and find a pronounced anisotropy in the thermopower 
for thermal gradients applied parallel and perpendicular to the Ru lamellae.
To clarify its origin, we combine thermopower measurements with electronic-structure-based transport calculations and finite-element modeling of the local strain distribution generated by the thermal-expansion mismatch between \sro and Ru. 
The calculations show that uniaxial compression of \sro produces a sizable thermopower anisotropy through modification of the Fermi surfaces. 
When an idealized spatially inhomogeneous strain field around the Ru lamellae is included, 
the observed sign and qualitative magnitude of the anisotropy are reproduced more closely than in a simple unstrained two-component model. 
These results are consistent with a substantial contribution from strain-induced changes in the \sro electronic structure near the interfaces, 
although contributions from interfacial phase mixing and three-dimensional transport connectivity cannot be excluded.
\end{abstract}

\maketitle

\section{introduction}
Sr$_2$RuO$_4$ is a layered perovskite superconductor with a bulk transition temperature of
$T_{\mathrm c}\simeq 1.5$~K and has long served as a prototypical platform for studying
unconventional superconductivity \cite{Maeno1994,RiceSigrist1995,Mackenzie2017,Maeno2024,Maeno2024JPSJ}.
Its normal-state electronic structure is well characterized: the low-energy states are derived
from the Ru $t_{2g}$ orbitals and form three Fermi-surface sheets, namely the quasi-one-dimensional
$\alpha$ and $\beta$ sheets and the quasi-two-dimensional $\gamma$ sheet \cite{Mackenzie1996,Damascelli2000,Bergemann2000}. 
Because superconductivity
emerges from this clean, multiband, and strongly anisotropic metal, Sr$_2$RuO$_4$ provides an
opportunity to investigate how orbital character, Fermi-surface topology, and symmetry-breaking
perturbations shape an unconventional superconducting state. At the same time, the symmetry of the
superconducting order parameter remains under active debate \cite{Pustogow2019,Ishida2020,Petsch2020,Benhabib2021,Ghosh2021,Grinenko2021},
which makes it particularly important to identify probes that are directly sensitive to the underlying
electronic structure.

When \sro crystals are grown under Ru-rich conditions, a \sroru eutectic system is obtained,
in which metallic Ru lamellae are embedded in a Sr$_2$RuO$_4$ single crystal \cite{Maeno1998}.
This composite exhibits a superconducting onset near 3~K, 
the so-called 3-K phase, 
substantially higher than the bulk $T_{\mathrm c}$ of pure \sro \cite{Maeno1998,Ando1999,Wada2002,Yaguchi2003}.
It has been shown that this enhanced superconductivity is not a bulk property of the
entire crystal but instead develops only within a limited volume near the interface between \sro and Ru,
indicating that the 3-K phase is fundamentally an interfacial phenomenon \cite{Mao2001,Yaguchi2003,Ying2009,Kittaka2009,Kittaka2009L,Kambara2009,Kaneyasu2010}.
Its microscopic origin, however, remains unresolved. Since the superconducting state of bulk
Sr$_2$RuO$_4$ is highly sensitive to lattice distortion and disorder \cite{Mackenzie1998}, 
several ingredients may be relevant in the eutectic system,
including local symmetry lowering at the interface, modification of the pairing interaction, and
lattice distortion induced by the Ru inclusions.
These intertwined effects make it difficult to isolate the role of the Ru inclusions in the enhancement of
$T_{\rm c}$ in the eutectic system.

An important clue comes from uniaxial-stress studies of pure Sr$_2$RuO$_4$ \cite{Kittaka2010,Hicks2014,Taniguchi2014,Steppke2017,Sunko2019,Jerzembeck2022,Jerzembeck2023}.
In-plane stress along the [100] direction strongly enhances $T_{\mathrm c}$ and drives a pronounced reconstruction
of the Fermi surface by shifting the van Hove singularity (vHs) of the $\gamma$ band toward the Fermi level,
eventually inducing a Lifshitz transition \cite{Steppke2017,Sunko2019}.
Because the in-plane strain breaks the tetragonal symmetry and lifts 
the equivalence between the vHs points at $(\pi,0)$ and $(0,\pi)$ \cite{Sunko2019}, 
it also generates a marked in-plane anisotropy in the electronic structure 
once the $\gamma$ sheet approaches the vHs point. 
In the eutectic system, the mismatch in thermal
contraction and elastic response between Ru and Sr$_2$RuO$_4$ naturally gives rise to a spatially
inhomogeneous strain field around the embedded Ru lamellae. 
If the local electronic structure near
the interface between \sro and Ru is tuned in a manner analogous to externally applied in-plane strain, 
such a strain field near the Ru lamellae may enhance the local $T_{\rm c}$ \cite{Steppke2017}.
In such a case, the highly anisotropic shapes of the embedded Ru lamellae may also lead to strong in-plane anisotropy.

To examine the anisotropic strain effect in the \sroru eutectic system,
here we investigate the in-plane thermopower in the normal state of the eutectic system for
heat currents applied parallel and perpendicular to the Ru lamellae,
and 
observe a clear anisotropy in the thermoelectric response. 
To identify its origin, we combine thermopower measurements with
first-principles-based transport calculations under uniform strain and finite-element modeling of the
local strain distribution generated by the Ru inclusions. 
We show that the experimentally observed
anisotropy is qualitatively explained by strain-induced modification of the electronic structure of
Sr$_2$RuO$_4$ on the Sr$_2$RuO$_4$ side of the interface. 
Our results establish thermopower as a sensitive
probe of interfacial strain in the eutectic system and provide a microscopic framework for discussing
the anomalous normal-state transport relevant to the 3-K phase.

\begin{figure}[t]
\begin{center}
\includegraphics[width=8.5cm]{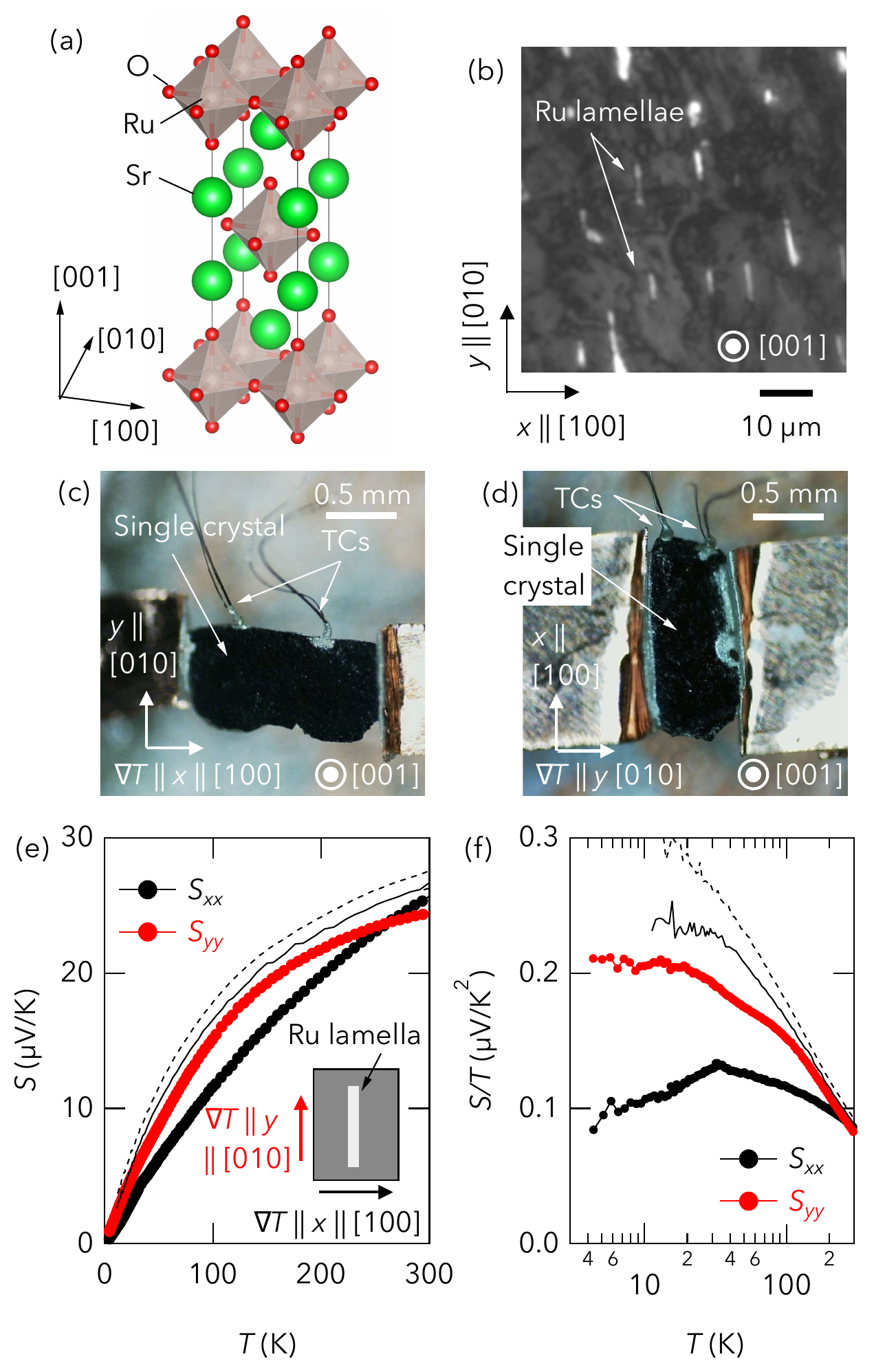}
\caption{
(a) Crystal structure of Sr$_2$RuO$_4$ visualized using VESTA \cite{Momma2011}.
(b) Photograph of the crystal surface of the \sroru eutectic sample. 
The bright regions indicate the embedded Ru lamellae. 
In the measured crystal, 
the Ru lamellae were oriented along the [100] (or equivalent [010]) direction of the \sro crystals.
The $x$ direction refers to the direction perpendicular to the long axis of the Ru lamellae, 
while the $y$ direction refers to the direction parallel to it,
as is also illustrated in the inset of (e).
(c,d) Photograph of the measurement setup for (c) $\nabla T || x$ and (d) $\nabla T || y$. 
Two thermocouples (TCs) are attached on the side of the single crystal to measure the temperature and the potential differences.
(e,f) Temperature dependence of (e) the measured thermopower $S$ and (f) $S/T$.
The black and red symbols indicate the data of the eutectic sample measured along the $x$ and $y$ directions, respectively.
The solid and dashed curves show the data of pure \sro single crystal without Ru inclusions measured along the [100] and [010] directions, respectively.
}
\label{fig:setup}
\end{center}
\end{figure}

\section{Methods}
Single crystals of the Sr$_2$RuO$_4$--Ru eutectic system were grown by a floating-zone method \cite{Mao2000MRB,Yaguchi2003}.
The crystal dimension was approximately $1.5\times0.6\times0.1~\mathrm{mm}^3$.
Figure~\ref{fig:setup}(a) depicts the unit cell of \sro and
Fig.~\ref{fig:setup}(b) shows a photograph of the in-plane crystal surface of the measured \sroru eutectic sample. 
The $c$ axis ([001] direction) is perpendicular to the imaged and measured surface.
The bright regions indicate the embedded Ru lamellae. 
Note that
the lamellae are assigned to Ru on the basis of the established \sroru eutectic growth protocol and the characteristic morphology reported in Refs. \onlinecite{Maeno1998,Ando1999,Wada2002,Yaguchi2003}, 
but we did not perform nanoscale chemical-state analysis on the measured sample.
Although the area shown in Fig.~\ref{fig:setup}(b) is a portion of the crystal, 
the Ru lamellae were predominantly aligned along the [100] (or equivalent [010]) direction in the whole crystal.
Note that the crystal axes were determined using a Laue camera.
The area fraction of the Ru lamellae in the crystal surface was about $2\%$,
which was determined using the binarized image.
Note that the distribution along the $c$-axis direction is unknown.
In this study, 
the $x$ and $y$ directions refer to the directions perpendicular and parallel to the long axis of the Ru lamellae, respectively,
and we measured the thermopower by applying the temperature gradient along the $x$ and $y$ directions as schematically shown in the inset of Fig.~\ref{fig:setup}(e).
Figures~\ref{fig:setup}(c) and \ref{fig:setup}(d) show the photographs of the thermopower measurement setup for $\nabla T || x$ and $\nabla T || y$, respectively.
In this study, thermopower was measured on one eutectic sample and one pure \sro sample; each specimen was rotated and remounted to obtain the two orthogonal in-plane configurations.
The in-plane thermopower was measured by a steady-state technique using manganin-constantan differential thermocouples in a closed-cycle refrigerator \cite{Kouda2022PRR,Ohsumi2024}.
A typical temperature gradient of $0.5~\mathrm{K/mm}$, adjusted depending on the bath temperature, was applied 
using a resistive heater. 
The thermoelectric voltage originating from the lead wires was subtracted from the measured signal.
The thermopower measurement system was calibrated using a NIST Standard Reference Material 3451 \cite{Beekman2024}.

We also performed first-principles calculations based on density functional theory (DFT) using Quantum ESPRESSO~\cite{Giannozzi2009JPCM,Giannozzi2009JPCM_v2,Giannozzi2009JPCM_v3}.
We employed projector-augmented-wave (PAW) pseudopotentials together with the Perdew--Burke--Ernzerhof (PBE) generalized gradient approximation (GGA) for the exchange-correlation functional.
The kinetic-energy cutoffs for the plane-wave basis set and the charge density were set to 70~Ry and 560~Ry, respectively, and a uniform $30 \times 30 \times 30$ $k$-point mesh was employed to ensure numerical convergence.
We carried out fully relativistic calculations including spin--orbit coupling.
From the resulting band energies $E_{n,\mathbf{k}}$, the density of states was evaluated as $D(\varepsilon)=\sum_{n,\mathbf{k}}\delta(\varepsilon-E_{n,\mathbf{k}})$ using the optimized tetrahedron method~\cite{Kawamura2014}, 
where $\delta$ denotes the Dirac delta function.
The Fermi surfaces were drawn by FermiSurfer \cite{Kawamura2019}.
The transport coefficients were calculated based on the linearized Boltzmann equations under the constant-relaxation-time approximation \cite{Madsen2006,Manako2024}.
The constant relaxation time of $10^{-14}$~s was used but it does not affect the thermopower.
A smoothed Fourier interpolation with a 20-times-denser $k$ grid was used for the transport calculation \cite{Pickett1988}.
Since a large number of mesh points are required at low temperatures, 
resulting in high computational costs, 
we have calculated the transport coefficients for temperatures above 100~K.
Accordingly, their comparison with experiment is qualitative and is not used to interpret the low-temperature regime.

\section{Results and Discussion}

Figures~\ref{fig:setup}(e) and \ref{fig:setup}(f) show the temperature dependence of the thermopower $S_{ii}$ and $S_{ii}/T$ measured for the $i$ direction ($i=x,y$) in the \sroru eutectic sample, respectively.
In these figures, we also plot the in-plane thermopower data of the pure \sro reference crystal as the solid and dashed curves, which represent the data measured along the $a$-axis ([100]) and $b$-axis ([010]) directions, respectively.
The overall behavior of the in-plane thermopower is similar to that observed in pure \sro crystals \cite{Yoshino1996,Xu2008,Yamanaka2022PRB,Daou2023,Otsuki2023,Nishinakayama2024},
but a notable anisotropy between the $x$ and $y$ directions is observed in the present \sroru eutectic system.
Note that a considerable sample dependence of the in-plane thermopower has been reported in earlier studies \cite{Yoshino1996,Xu2008,Yamanaka2022PRB}. 
On the other hand, since we measured a single sample along different directions, 
it is unlikely that the observed anisotropy was due to sample-dependent effects, 
as was the case in such previous studies.

In Figs.~\ref{fig:setup}(e) and \ref{fig:setup}(f),
the $a$-axis and $b$-axis data of the pure \sro crystal (solid and dashed curves) must be identical due to the tetragonal symmetry.
Their small difference provides an empirical estimate of systematic uncertainty introduced by remounting, thermocouple placement, and geometric misalignment.
The in-plane thermopower values of \sro at $T=300$ K in the earlier reports 
are listed 
in Table I.
Although there is some variation, data of pure \sro crystals show similar values including this work.
On the other hand, 
as is discussed later, 
the room-temperature thermopower values in the eutectic sample are smaller than that of the pure sample for both $x$ and $y$ directions.
Nevertheless, 
the observed anisotropy between the $x$ and $y$ directions in the eutectic crystal is considerably larger than that of the pure reference crystal between the $a$ and $b$ directions in the measured temperature range,
although the use of one specimen of each type limits conclusions regarding sample-to-sample reproducibility.

\begin{table}
\label{table:sdata}
\begin{center}
\caption{Comparison of the in-plane thermopower at $T=300$ K. }
\begin{tabular}{lrl} \hline
    &  $S_{}$ ($\mu$V/K) & Ref.\\ \hline
   Yoshino $et$ $al$., sample 1 & 28.5 & \cite{Yoshino1996} \\
   Yoshino $et$ $al$., sample 2 & 27.7 & \cite{Yoshino1996} \\ 
   Xu $et$ $al$., sample A & 27.5 & \cite{Xu2008} \\ 
   Xu $et$ $al$., sample B & 29.5 & \cite{Xu2008} \\  
   Xu $et$ $al$., sample C & 29 & \cite{Xu2008} \\ 
   Yamanaka $et$ $al$.  & 25 & \cite{Yamanaka2022PRB} \\ 
   Daou $et$ $al$.  & 27 & \cite{Daou2023} \\ 
   Pure sample ($a$ direction)  & 26.5 & This work \\ 
   Pure sample ($b$ direction)  & 27.5 & This work \\ 
   Eutectic sample ($x || [100]$ direction)  & 25.5 & This work \\ 
   Eutectic sample ($y || [010]$  direction)  & 24.5 & This work \\ \hline
\end{tabular}
\end{center}
\end{table}

\begin{figure}[t]
\begin{center}
\includegraphics[width=8.8cm]{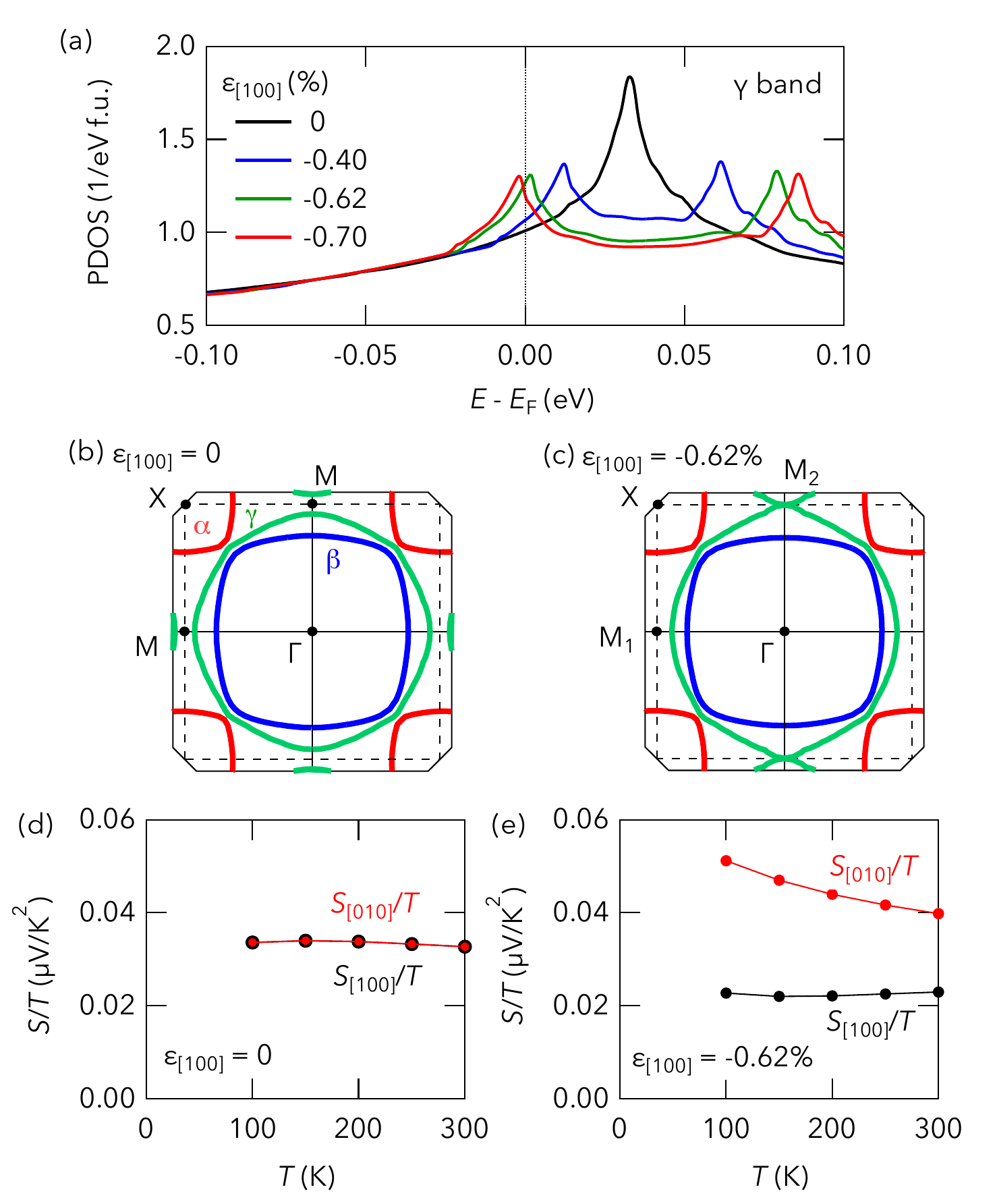}
\caption{
(a) 
Energy dependence of the calculated partial density of states (PDOS) from the $\gamma$ band of \sro for a set of uniform strain $\varepsilon_{[100]}$.
(b,c) Calculated Fermi surfaces under (b) $\varepsilon_{[100]}=0$ and (c) $-0.62\%$.
(d,e) Calculated thermopower divided by temperature $S/T$ for (d) $\varepsilon_{[100]}=0$ and (e) $-0.62\%$.
The black and red symbols indicate the thermopower calculated for the [100] and [010] directions, respectively.
}
\label{fig:calc}
\end{center}
\end{figure}

To examine the strain effect on the observed anisotropic thermopower in the \sroru eutectic system, 
we then performed the first-principles calculations under uniaxial pressure.
For the strain along the [100] direction $\varepsilon_{[100]}$, 
other strains are determined using the Poisson's ratios as
$\varepsilon_{[010]} = -\nu_{xy}\varepsilon_{[100]}$ and 
$\varepsilon_{[001]} = -\nu_{xz}\varepsilon_{[100]}$,
where 
$\nu_{xy} = 0.39$ and $\nu_{xz}=0.21$ \cite{Paglione2002PRB}.
Then, calculations were performed for the distorted lattices
with the lattice parameters, $a$, $b$, and $c$, defined as $a=a_0(1+\varepsilon_{[100]})$, $b=a_0(1+\varepsilon_{[010]})$, and 
$c=c_0(1+\varepsilon_{[001]})$,
where $a_0$ and $c_0$ are the lattice parameters of \sro measured at $T=15$~K \cite{Chmaissem1998}.

The calculation results under uniform strain are summarized in Fig.~\ref{fig:calc}.
Figure~\ref{fig:calc}(a) shows the partial density of states (PDOS) for the $\gamma$ band
calculated with $\varepsilon_{[100]}=0$, $-0.4\%$, $-0.62\%$, and $-0.7\%$. 
As the strain increases, the vHs point splits, and one of the resulting points approaches $E_{\rm F}$.
The critical value of $\varepsilon_{[100]}$ is approximately $-0.62\%$, which is close to the literature value \cite{Steppke2017}.
The Fermi surfaces for unstrained and strained lattices are shown in Figs.~\ref{fig:calc}(b) and \ref{fig:calc}(c), respectively.
In the strained case with $\varepsilon_{[100]} = -0.62\%$, 
the $\gamma$ sheets connect to the adjacent sheet at the $M_2$ point.

Figures~\ref{fig:calc}(d) and \ref{fig:calc}(e) show the calculated thermopower divided by temperature $S/T$ for the 
unstrained ($\varepsilon_{[100]} = 0$) and strained lattices ($\varepsilon_{[100]} = -0.62\%$), respectively.
We calculated the thermopower for the compressed $[100]$ direction, $S_{[100]}$, and that for the tensiled $[010]$ direction, $S_{[010]}$.
In the unstrained case [Fig.~\ref{fig:calc}(d)], there is no in-plane anisotropy of thermopower owing to tetragonal symmetry.
Note that correlation effects of the 4$d$ electrons are significant in \sro and 
the calculation results within the local-density approximation are quantitatively different from the experimental values of $S/T$ \cite{Mravlje2016}.
In the strained case [Fig.~\ref{fig:calc}(e)], on the other hand, 
a prominent thermopower anisotropy is found between $S_{[100]}$ and $S_{[010]}$.
The enhanced thermopower in the strained \sro crystal has been calculated \cite{Herman2019}, and 
here we focus on the in-plane thermopower anisotropy.
As observed in the present \sroru eutectic system [Fig.~\ref{fig:setup}(f)], 
the in-plane thermopower anisotropy calculated in the strained crystal 
implies that the strain effect may be significant for the eutectic sample.

As shown in Figs.~\ref{fig:setup}(e) and ~\ref{fig:setup}(f), 
the thermopower measured along the $y$ direction, where the heat current is applied parallel to the Ru lamellae, 
is relatively enhanced compared with that measured along the $x$ direction, 
where the heat current is applied perpendicular to the Ru lamellae. 
Along with the calculation results under strain ($S_{[100]} < S_{[010]}$),
the experimental trend ($S_{xx} < S_{yy}$) suggests that 
the precipitation of Ru lamellae induces a compressive strain along the $x$ direction ($\varepsilon_{[100]}<0$) in the surrounding \sro
owing to the anisotropic shape of the Ru lamellae elongated along the $y$ direction.
Note that although
the experimental data show the opposite trend ($S_{xx} > S_{yy}$) near room temperature,
the origin of which is unclear at present,
an overall trend of $S_{xx} < S_{yy}$ in the wide temperature range [Fig.~\ref{fig:setup}(f)]
implies such a compressive strain along the $x$ direction.

These results are qualitatively explained by the thermally induced strain model 
as discussed in Ref.~\cite{Ando1999}.
The linear thermal expansions from 2000 K to 3 K are $-1.87\%$ and $-2.55\%$ for
the Ru metal and the in-plane direction of \sro, respectively \cite{Chmaissem1998,Finkel1971},
indicating a larger thermal contraction of \sro along the in-plane direction compared with that of Ru upon cooling from the eutectic growth temperature to low temperature. 
In the absence of interfacial constraint, the $a$-axis lattice parameter of \sro would shrink more than that of Ru. 
At the interface, however, this contraction is partially suppressed by the less-contracting Ru inclusion, 
forcing the adjacent Sr$_2$RuO$_4$ region to retain a larger in-plane lattice parameter than in the freely contracted state. 
As a result, 
a positive strain is induced in the $y$ direction (the direction of interface extension) of \sro ($\varepsilon_{[010]}>0$), 
while a negative strain occurs in the $x$ direction ($\varepsilon_{[100]}<0$).
This strain effect due to the Ru lamellae may cause the observed anisotropic thermopower.

\begin{figure}[t]
\begin{center}
\includegraphics[width=8.7cm]{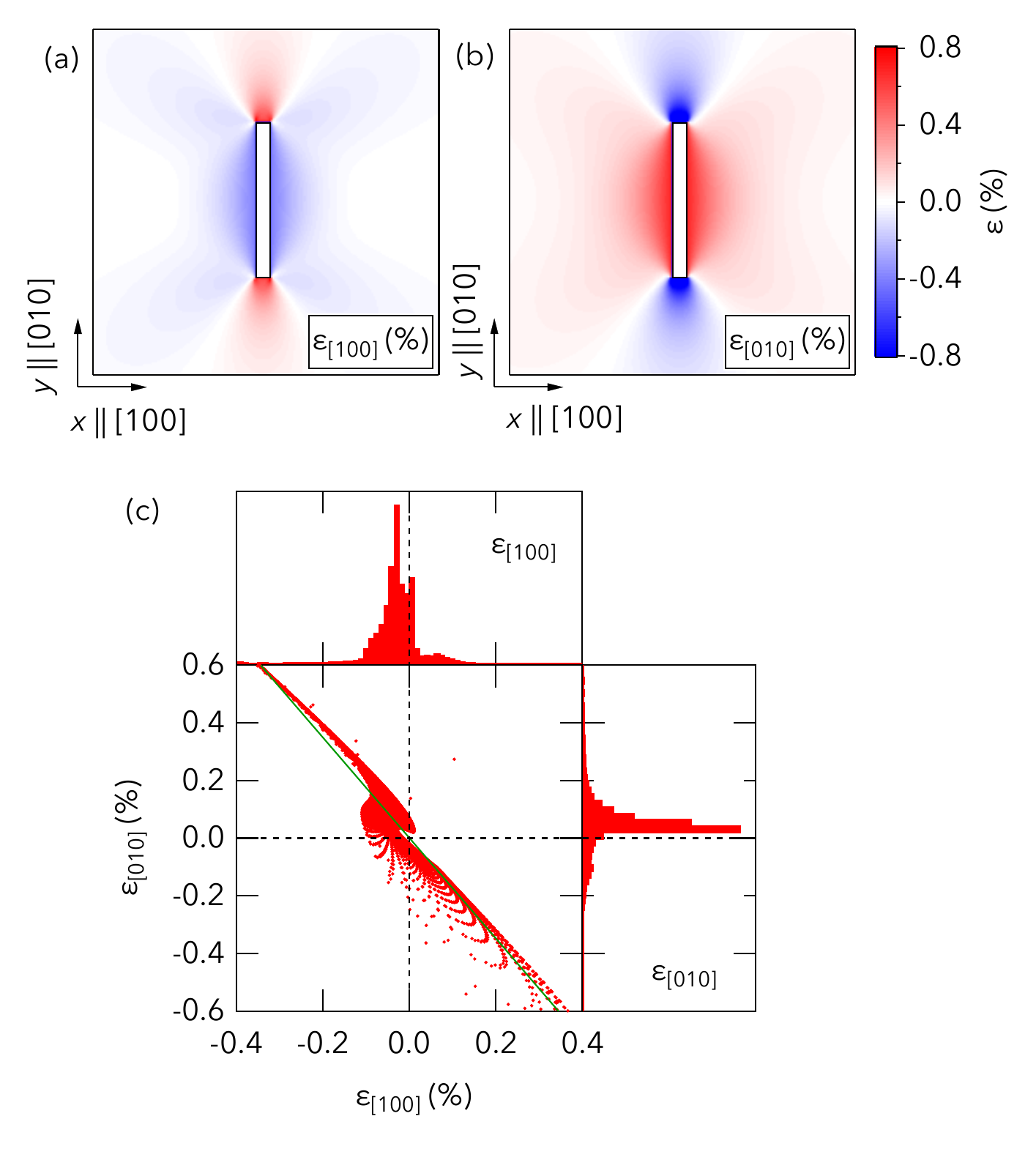}
\caption{
(a,b) Contour plot of (a) $\varepsilon_{[100]}$ and (b) $\varepsilon_{[010]}$ calculated with the finite element method (FEM). 
A rectangular Ru inclusion was embedded in the center of the \sro matrix.
(c) Distribution of $\varepsilon_{[100]}$ and $\varepsilon_{[010]}$ in (a,b) and the corresponding linear fit (solid line). Histograms of $\varepsilon_{[100]}$ and $\varepsilon_{[010]}$ are also plotted.
}
\label{fig:fem}
\end{center}
\end{figure}

\begin{figure}[t]
\begin{center}
\includegraphics[width=8.7cm]{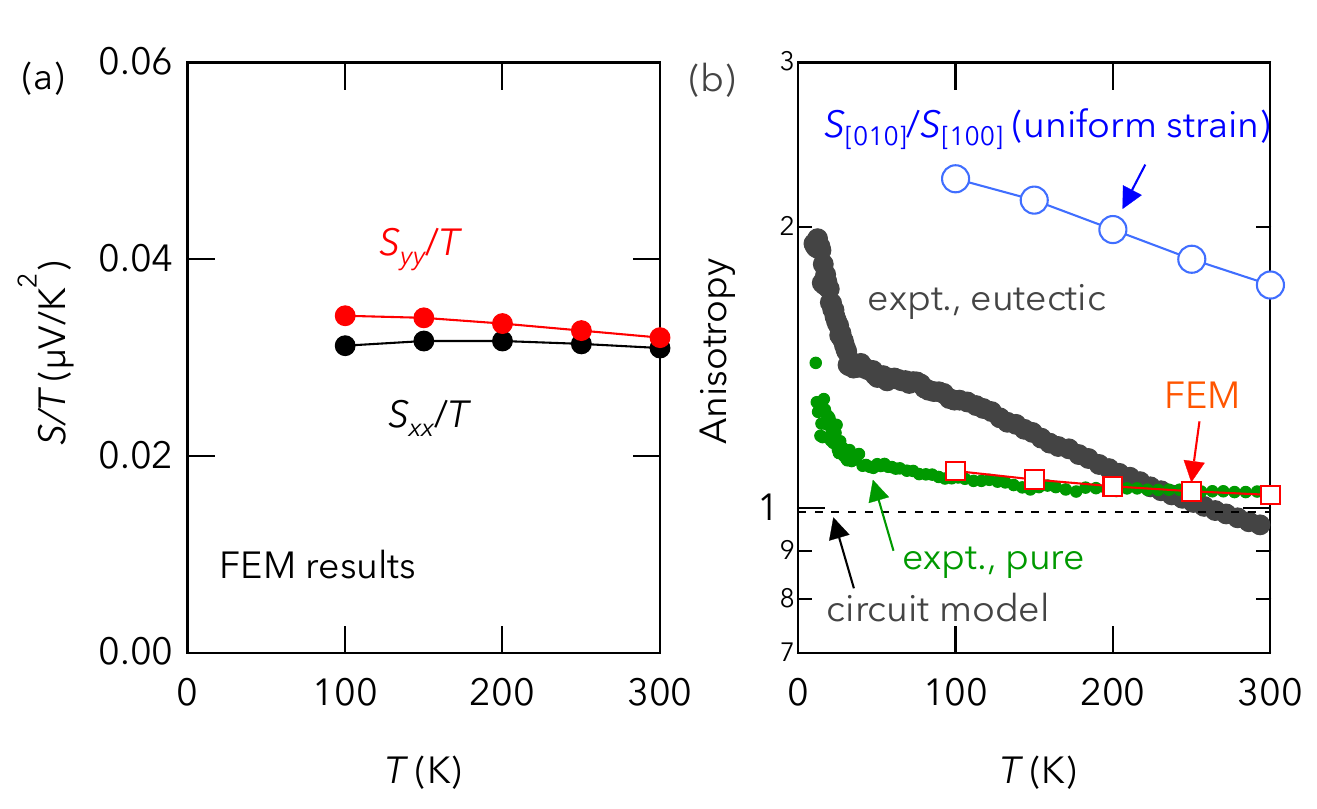}
\caption{
(a) Temperature dependence of $S/T$ calculated for the $x$ and $y$ directions in the FEM scheme. 
(b) The thermopower anisotropy. 
The experimental results are shown as the black symbols for the eutectic sample and as the green symbols for the pure \sro sample.
The calculation results under the uniform strain and the FEM scheme are shown in the blue and red symbols, respectively.
The dashed line indicates the anisotropy estimated from a two-component circuit model
consisting of the unstrained \sro crystal and Ru inclusion.
}
\label{fig:femS}
\end{center}
\end{figure}

In calculations under uniaxial pressure, 
the strain field is assumed to be uniform throughout the sample. 
In the \sroru eutectic system, however, 
local strain may be generated around the Ru inclusions. 
Therefore, to estimate the thermopower in such an inhomogeneous strain field, 
we evaluate a strain distribution using the finite-element method (FEM).
In the present program, 
we fixed the size of the embedded Ru lamellae and calculated the strain fields, $\varepsilon_{[100]}, \varepsilon_{[010]}, \varepsilon_{[001]}$, for the surrounding \sro 
using 
the elastic constants of \sro \cite{Paglione2002PRB} and the differential linear thermal expansion coefficient between Sr$_2$RuO$_4$ and Ru \cite{Ando1999}. 
Details are described in the Appendix section.

Figures~\ref{fig:fem}(a) and \ref{fig:fem}(b) show the contour plots of $\varepsilon_{[100]}$ and $\varepsilon_{[010]}$ calculated with the FEM scheme, respectively.
In this calculation, we set a rectangular-shaped Ru inclusion with an area fraction of $2\%$,
which is approximately the experimental value.
We also set the aspect ratio of the Ru inclusion between the $x$- and $y$-directions to $1:10$, 
which is the average value obtained from the experiment.
As expected, the regions near the interfaces are strongly strained, 
and many regions are indeed compressed along the $x$ direction ($\varepsilon_{[100]}<0$) due to the anisotropic shape of the Ru inclusion.
The distribution of $\varepsilon_{[100]}$ and $\varepsilon_{[010]}$ is plotted in Fig.~\ref{fig:fem}(c).
Note that the average slope in this $\varepsilon_{[100]}$-$\varepsilon_{[010]}$ plot is significantly 
different from the Poisson's ratio of $\nu_{xy} = 0.39$, 
since this simulation does not involve a uniaxial stress state.
Based on this strain distribution, we then performed first-principles and transport calculations. 
We evaluated the thermopower and electrical conductivity for each cell, 
and finally calculated the total thermopower using a circuit model as described in the Appendix section.
Note that the effective thermopower is estimated using a simplified series/parallel circuit model, which neglects the interface electrical/thermal resistance and thermal-conductivity distribution.
Figure~\ref{fig:femS}(a) shows the temperature dependence of $S/T$ calculated for the $x$ and $y$ directions in the FEM scheme. 
The anisotropic thermopower is indeed obtained in the FEM calculation and the anisotropy ratio $S_{yy}/S_{xx}$ is smaller than
that for the uniform-strain case, since the strains are only enhanced near the Ru inclusions.
We have also examined the sensitivity of the FEM results to the used parameters, and qualitatively similar results are obtained (Appendix).

Figure~\ref{fig:femS}(b) compares the thermopower anisotropy obtained in this study.
We also calculate the thermopower anisotropy by assuming a simple two-component circuit model consisting of the unstrained \sro crystal and Ru inclusion (Appendix).
The resulting thermopower anisotropy ratio was almost unity as shown with the dashed line in Fig.~\ref{fig:femS}(b) and the deviation was less than 1\%, which is too small to explain the observed anisotropy.
The anisotropy calculated with the uniform strain of $\varepsilon_{[100]}=-0.62\%$ may be overestimated.
In contrast, 
although the temperature dependence is not reproduced well,
a moderate anisotropy is obtained even in the FEM calculation,
indicating that the local strain effect is essential for the \sroru eutectic system.
Correlation effects may be crucial to examine the thermopower anisotropy more quantitatively \cite{Buhmann2013,Mravlje2016}.
In addition, it is necessary to incorporate the non-Fermi liquid effects that arise as the system approaches the vHs point \cite{Barber2018,Herman2019,Georges2021}.
Also, the discrepancy can arise from various factors,
such as the constant-relaxation-time approximation, 
the use of a static low-temperature strain map for comparison, 
uncertainty in the temperature-dependent thermal-expansion mismatch, 
unknown $c$-axis morphology and inclusion fraction, 
the rigid-body inclusion, interface scattering, neglected thermal-conductivity contrast, possible oxidized interfaces, and experimental contact/misalignment uncertainty. 
These factors can affect both the magnitude and temperature dependence.

Here we comment on the percolative shunting effect of the embedded lamellae.
The percolation threshold is typically 10-25\% \cite{Shklovskii},
which is much larger than the present area fraction of Ru lamellae.
Indeed, Maeno $et$ $al$. compared samples cut from the Ru-free sheath and
the \sroru eutectic region of the same crystal rod and 
reported that their normal-state in-plane resistivities were nearly identical 
up to room temperature \cite{Maeno1998}. 
This observation indicates that Ru lamellae do not necessarily produce appreciable in-plane electrical shunting in typical eutectic crystals. 
In contrast, the out-of-plane resistivity of the eutectic system remains 
metallic up to 300 K 
because of composite current paths involving the Ru lamellae \cite{Ando1999},
demonstrating that the influence of the inclusions is strongly direction- 
and connectivity-dependent. 
Thus a sample-specific shunting contribution cannot be completely excluded.

The two-dimensional FEM represents an effective \(ab\)-plane cross-section of the eutectic structure. This treatment is motivated by the in-plane measurement geometry and by previous resistivity studies showing that the normal-state \(\rho_{ab}\) is only weakly affected by the Ru inclusions \cite{Maeno1998}.
However, the metallic \(c\)-axis transport of the eutectic system demonstrates the existence of three-dimensional composite conduction paths \cite{Ando1999}. The present model therefore neglects \(c\)-axis strain relaxation, finite lamella terminations, and out-of-plane current and heat-flow redistribution. 
The calculated strain distribution should consequently be interpreted as a qualitative description of the leading in-plane anisotropy rather than a quantitatively complete three-dimensional strain field.

Nevertheless, the strain effect has recently been examined in various quantum materials and correlated electron systems to modify their electronic structures \cite{Aramberri2017,Kawasugi2019}.
Thus, the present study suggests that eutectic crystal growth may be an effective way of inducing strain without the need for specialized experimental equipment. 
It has been proposed that eutectic growth is also effective for enhancing the efficiency of the thermoelectric effect, as interfacial scattering suppresses the lattice thermal conductivity \cite{Sootsman2010}.
The present results indicate that, by controlling the eutectic phase, it may be possible not only to suppress thermal conductivity but also to increase the thermopower toward efficient thermoelectrics.

\section{Summary}
We report anisotropic normal-state thermopower in the \sroru eutectic system. 
The thermopower differs markedly for thermal gradients applied parallel and perpendicular to the Ru lamellae. 
Within the assumptions of the present models, simple geometrical mixing of unstrained \sro and Ru produces a much smaller anisotropy than observed, whereas inclusion of the calculated local strain field yields the correct sign and a closer qualitative magnitude. This comparison supports local strain as an important factor for the observed thermopower anisotropy in the eutectic sample. 
However, on the quantitative discussion, 
the present data do not uniquely separate strain from other factors such as interfacial phase chemistry or three-dimensional percolative transport.
More microscopic experiments will be required to examine the anisotropic normal-state properties of the eutectic crystals.

\section*{Acknowledgements}
We thank R. Kurihara for discussions and A. Ito for assistance with sample preparation.
This work was partly supported by JSPS KAKENHI Grant No. 22H01166 and No. 24K06945, 
a research grant from Murata Science and Education Foundation (M25AN115),
and Research Foundation for the Electrotechnology of Chubu (REFEC, No. R-07203).

\section*{Data availability}
The data are available from the authors upon reasonable request.

\section*{APPENDIX}

\subsection{In-plane anisotropy observed in the pure \sro crystal}

\begin{figure}[t]
\begin{center}
\includegraphics[width=8.6cm]{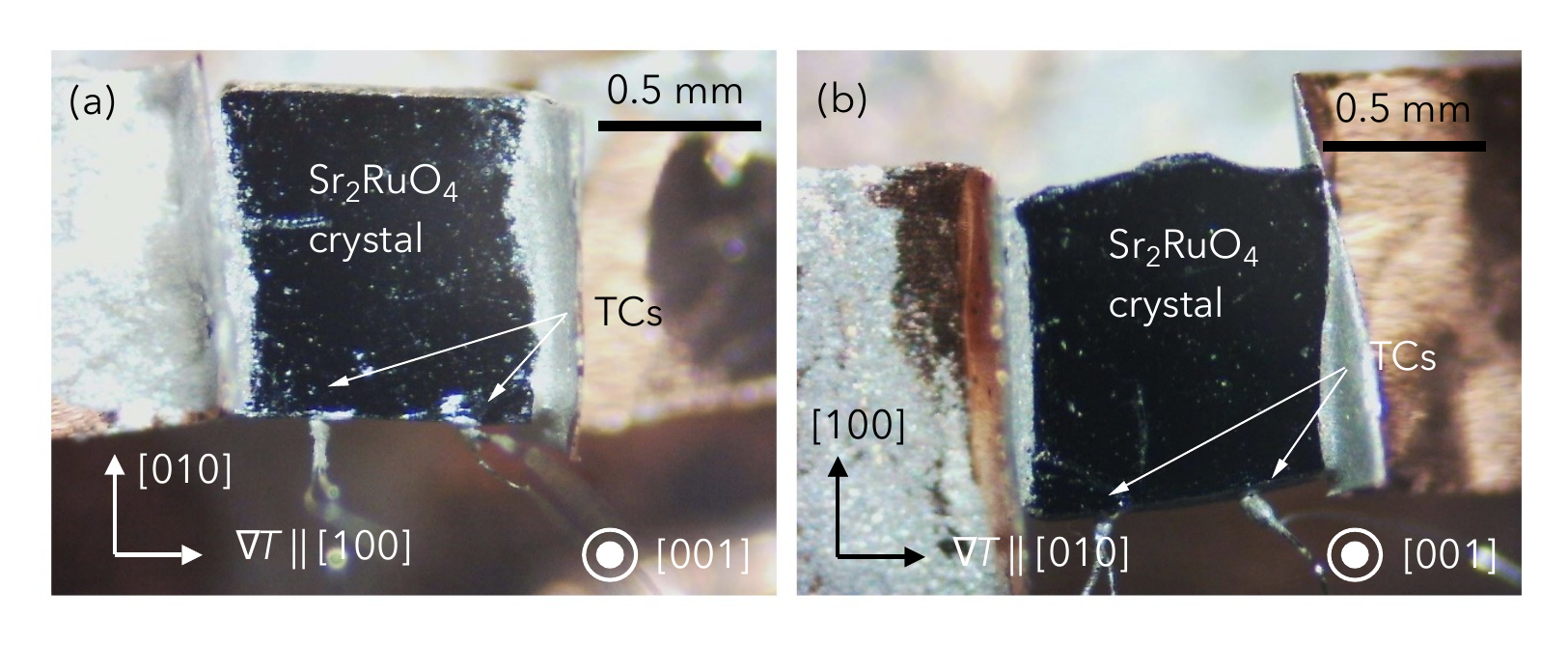}
\caption{
Photograph of the setup to measure the thermopower of the pure \sro crystal for (a) [100] ($a$-axis) and (b) [010] ($b$-axis) directions.}
\label{figA:setup}
\end{center}
\end{figure}

We examined the in-plane thermopower anisotropy in a pure \sro single crystal without Ru inclusions.
Here we measured the $a$-axis thermopower $S_{aa}$ and the $b$-axis thermopower $S_{bb}$, 
which must be identical to the equivalent $a$-axis thermopower by tetragonal symmetry, 
by rotating a single sample. 
Figures~\ref{figA:setup}(a) and ~\ref{figA:setup}(b) show the experimental setup 
to measure $S_{aa}$ and $S_{bb}$ of the pure sample, respectively. 
The solid and dashed curves in Fig~\ref{fig:setup}(e) show the temperature dependence 
of $S_{aa}$ and $S_{bb}$, respectively, and
the thermopower anisotropy in the pure and eutectic samples is compared in
Fig.~\ref{fig:femS}(b).
The in-plane thermopower anisotropy observed in the pure sample, $S_{bb}/S_{aa}$, may originate from 
unavoidable experimental uncertainties such as wiring misalignment.
On the other hand, the anisotropy in the eutectic sample, $S_{yy}/S_{xx}$, is significantly larger than 
that of the pure sample, indicating an intrinsic anisotropy in the eutectic system.

\subsection{Finite Element Method}

Finite element analysis was performed by considering the thermal strain arising from the difference 
between the linear thermal expansion coefficients between \sro and Ru.
We set a rectangular-shaped Ru inclusion with an area fraction of $2\%$ in the \sro matrix.
The aspect ratio of the Ru inclusion between the $x$- and $y$-directions was set to $1:10$.
The area fraction and the aspect ratio are approximately the experimentally averaged values.
The Ru inclusion was treated as a rigid body and 
the surrounding \sro was then deformed owing to the difference in the thermal expansion.
The linear thermal expansions from 2000 K to 3 K are $-1.87\%$ and $-2.55\%$ for
the Ru metal and the in-plane direction of \sro, respectively \cite{Chmaissem1998,Finkel1971},
yielding a differential linear thermal expansion coefficient of $-0.68\%$ for \sro.

The Sr$_2$RuO$_4$ matrix was treated as an anisotropic elastic medium in the
in-plane approximation, with the stiffness matrix
\begin{equation}
\mathbf{C} =
\begin{pmatrix}
C_{11} & C_{12} & 0 \\
C_{12} & C_{11} & 0 \\
0 & 0 & C_{66}
\end{pmatrix},
\end{equation}
where \(C_{11}=2.32\times 10^{11}\,\mathrm{Pa}\),
\(C_{12}=1.06\times 10^{11}\,\mathrm{Pa}\), and
\(C_{66}=0.612\times 10^{11}\,\mathrm{Pa}\) \cite{Paglione2002PRB}.
The stress $\boldsymbol{\sigma}$ is then related to the elastic strain $\boldsymbol{\varepsilon}$ 
as
\begin{equation}
\boldsymbol{\sigma}
=
\mathbf{C}
\left(
\boldsymbol{\varepsilon}
-
\boldsymbol{\varepsilon}_{\rm th}
\right),
\end{equation}
where \(\boldsymbol{\varepsilon}_{\rm th}\) is the thermal strain given as
\begin{equation}
\boldsymbol{\varepsilon}_{\rm th}
=
\lambda
\begin{pmatrix}
1 \\ 1 \\ 0
\end{pmatrix}
\end{equation}
with the differential linear thermal expansion coefficient $\lambda = -0.0068$ as noted above.
According to the principle of virtual work, 
the equilibrium condition in the absence of external mechanical loads is written as
an area integral over the two-dimensional element domain $\Omega$ of 
\begin{equation}
\int_{\Omega}
\delta \boldsymbol{\varepsilon}^{\mathsf{T}}
\boldsymbol{\sigma}
\,d\Omega
= 0,
\end{equation}
where $\delta \boldsymbol{\varepsilon}$ is the virtual strain \cite{Deeks2002}.
Substituting the elastic strain, we obtain 
\begin{equation}
\int_{\Omega}
\delta \boldsymbol{\varepsilon}^{\mathsf{T}}
\mathbf{C}
\boldsymbol{\varepsilon}
\,d\Omega
=
\int_{\Omega}
\delta \boldsymbol{\varepsilon}^{\mathsf{T}}
\mathbf{C}
\boldsymbol{\varepsilon}_{\rm th}
\,d\Omega .
\end{equation}

Within each finite element, the strain and virtual strain are expressed in
terms of the nodal displacement vector \(\mathbf{u}_{\rm e}\) and the virtual nodal
displacement vector \(\delta\mathbf{u}_{\rm e}\) as
\begin{equation}
\boldsymbol{\varepsilon}
=
\mathbf{B}\mathbf{u}_{\rm e},
\qquad
\delta\boldsymbol{\varepsilon}
=
\mathbf{B}\delta\mathbf{u}_{\rm e} ,
\end{equation}
where \(\mathbf{B}\) is the strain-displacement matrix. By substitution,
we obtain the element equation of
\begin{equation}
\mathbf{K}_{\rm e}
\mathbf{u}_{\rm e}
=
\mathbf{f}_{{\rm th,e}},
\end{equation}
where the element stiffness matrix $\mathbf{K}_{\rm e}$ 
and the thermal nodal force $\mathbf{f}_{{\rm th,e}}$ are respectively
defined as
\begin{equation}
\mathbf{K}_{\rm e}
=
\int_{\Omega_{\rm e}}
\mathbf{B}^{\mathsf{T}}
\mathbf{C}
\mathbf{B}
\,d\Omega ,
\end{equation}
and
\begin{equation}
\mathbf{f}_{{\rm th,e}}
=
\int_{\Omega_{\rm e}}
\mathbf{B}^{\mathsf{T}}
\mathbf{C}
\boldsymbol{\varepsilon}_{\rm th}
\,d\Omega.
\end{equation}
${\Omega_{\rm e}}$ denotes the area occupied by a single finite element. 
After assembling all elements, the global finite-element equation becomes
\begin{equation}
\mathbf{K}\mathbf{u}
=
\mathbf{f}_{\rm th},
\label{eq:global}
\end{equation}
where \(\mathbf{K}\) is the global stiffness matrix and
\(\mathbf{f}_{\rm th}\) is the equivalent nodal force.
The displacement field  \(\mathbf{u}\) is then obtained by solving Eq.~(\ref{eq:global}).
A far-field displacement boundary condition was imposed on the outer boundary of the computational domain. 
The mesh grid was set to $200\times200$.

For each deformed cell, 
we evaluated the strain components $\varepsilon_{[100]}$ and $\varepsilon_{[010]}$ at low temperature (3~K) 
caused by Ru inclusions 
based on the difference in lattice lengths between samples with and without Ru inclusions.
Because each cell is generally distorted into a parallelogram after deformation, 
the lengths of opposite sides are not necessarily identical.
Therefore, the lattice length along the $x$ direction is 
defined as the average of the top and bottom side lengths of each cell.
Similarly, 
the lattice length along the $y$ direction is also defined
as the average of the left and right side lengths.
The strain along the $x$ direction, $\varepsilon_{[100]}$, which is induced by the Ru inclusion,
is obtained for each cell as
\[
\varepsilon_{[100]}
=
\frac{a_x-a_0}{a_0},
\]
where $a_x$ and $a_0$ are the low-temperature $x$-direction lattice lengths thermally deformed with and without Ru inclusion, respectively.
The strain along the $y$ direction, $\varepsilon_{[010]}$, is similarly obtained as 
\[
\varepsilon_{[010]}
=
\frac{a_y-a_0}{a_0},
\]
where $a_y$ is the low-temperature $y$-direction lattice length thermally deformed with Ru inclusion.

\subsection{Thermopower calculation using the FEM results}

\begin{figure}[t]
\begin{center}
\includegraphics[width=8.6cm]{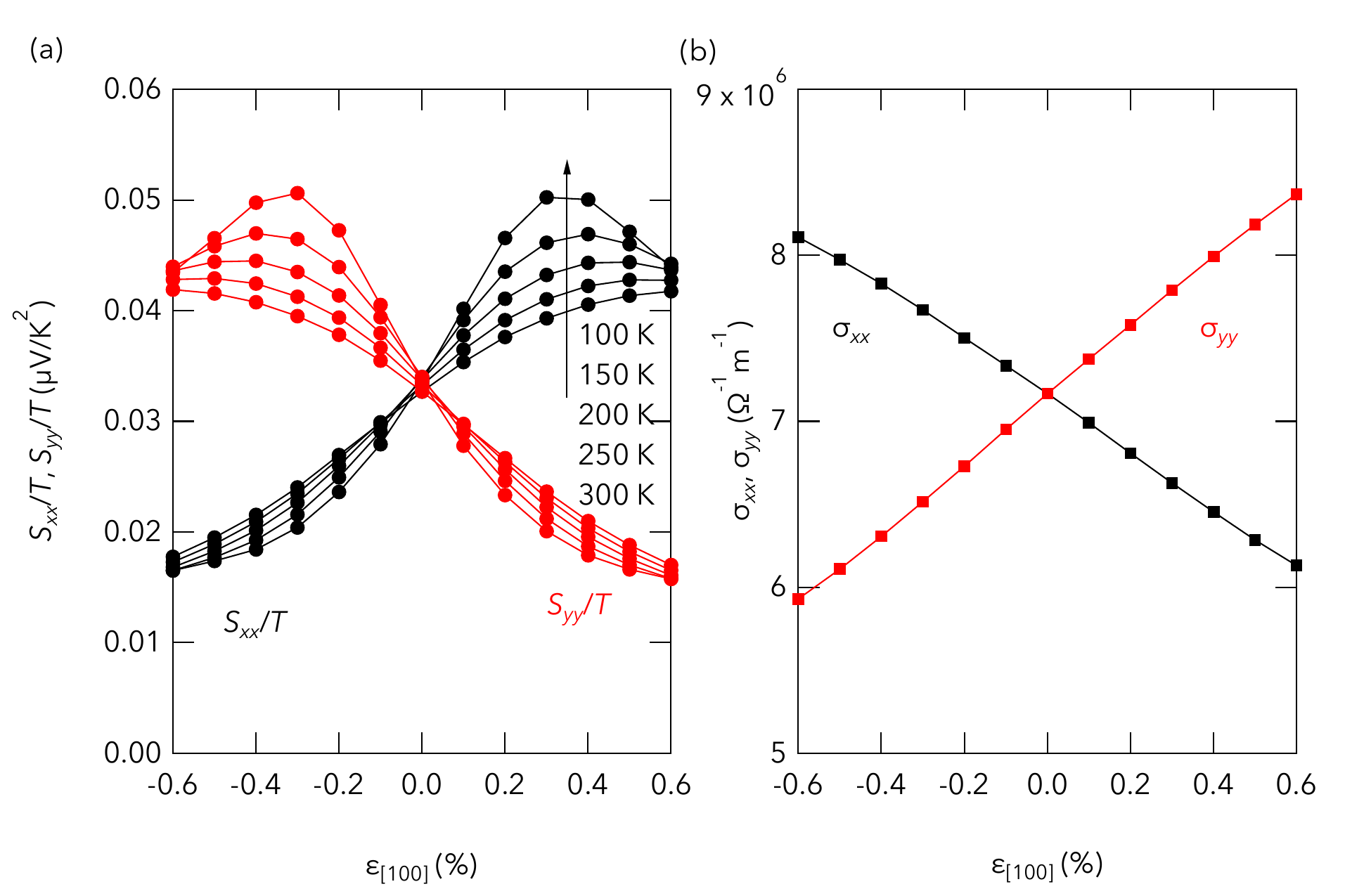}
\caption{
Transport coefficients [(a) thermopower divided by temperature, $S_{ii}/T$ ($i=x,y$), and (b)
electrical conductivity, $\sigma_{ii}$] 
calculated for several values of $\varepsilon_{[100]}$.
The electrical conductivity are almost temperature-independent since the constant relaxation time approximation is employed.
}
\label{figA:trans}
\end{center}
\end{figure}

\begin{figure*}[t]
\begin{center}
\includegraphics[width=16cm]{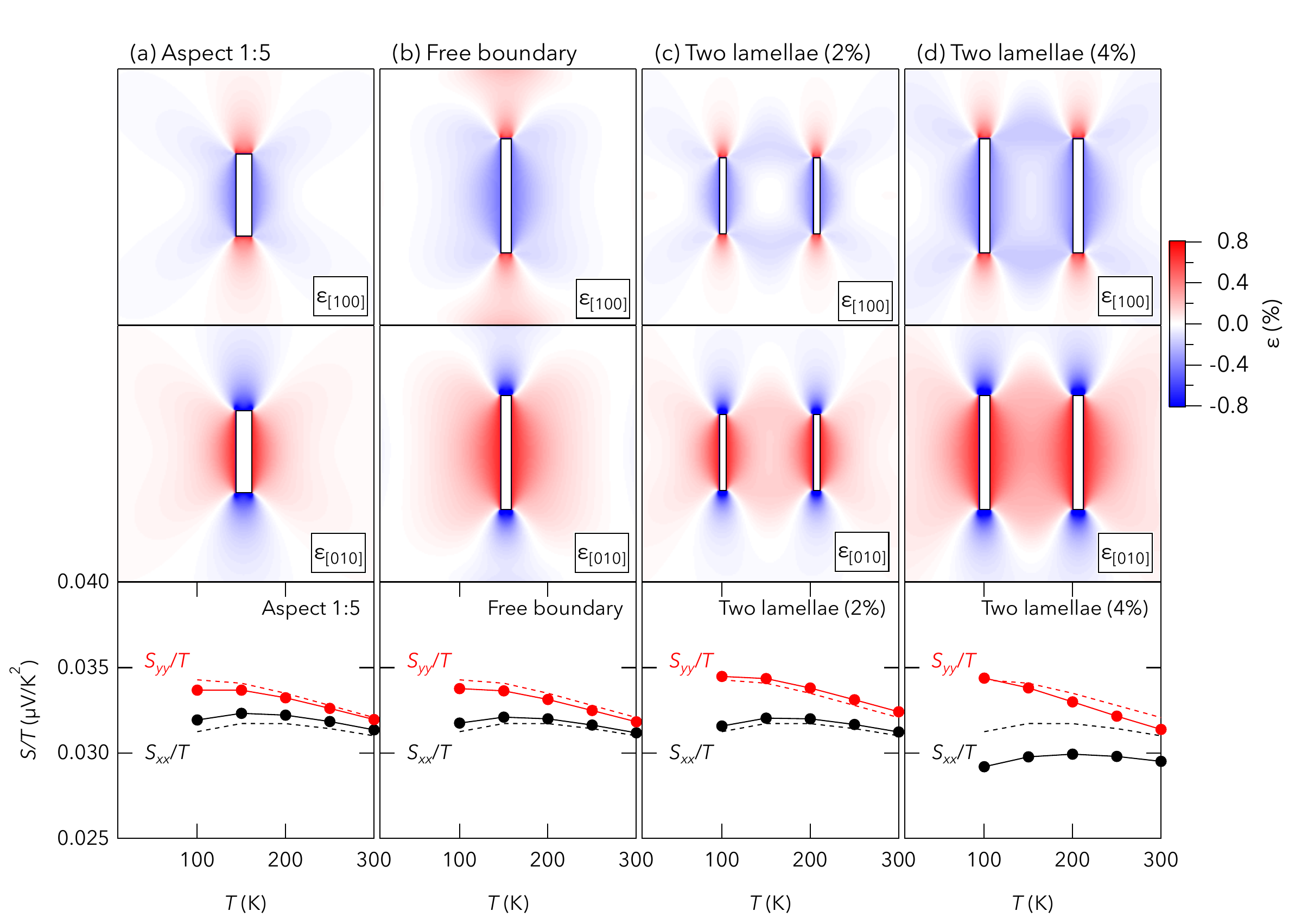}
\caption{
Sensitivity test of the anisotropic thermopower for various FEM conditions.
The calculation conditions are listed in Table II.
The top and middle panels show the contour plot of $\varepsilon_{[100]}$ and $\varepsilon_{[010]}$, respectively,
The bottom panels show the temperature dependence of 
$S/T$ calculated for the $x$ (black) and $y$ (red) directions.
In each panel, the dashed curves represent the calculated data with the common condition
noted in the main text (i.e. area fraction of 2\%, aspect ratio of 1:10, far-field boundary condition, and one lamella).
}
\label{figA:femall}
\end{center}
\end{figure*}

\begin{table*}
\label{table:conditions}
\begin{center}
\caption{Various conditions for the FEM calculation.}
\begin{tabular}{lrrrrr} \hline
    &  Area fraction of Ru inclusions & Aspect ratio & Boundary condition& No. of lamellae\\ \hline
   Fig.~\ref{fig:fem} (main text) & 2\% & 1:10 & far-field & 1 \\
   Fig.~\ref{figA:femall}(a) (appendix) & 2\% & 1:5 & far-field & 1 \\ 
   Fig.~\ref{figA:femall}(b) (appendix) & 2\% & 1:10 & free & 1 \\ 
   Fig.~\ref{figA:femall}(c) (appendix) & 2\% & 1:10 & far-field & 2 \\ 
   Fig.~\ref{figA:femall}(d) (appendix) & 4\% & 1:10 & far-field & 2 \\    
   \hline
\end{tabular}
\end{center}
\end{table*}

\begin{figure}[t]
\begin{center}
\includegraphics[width=8.6cm]{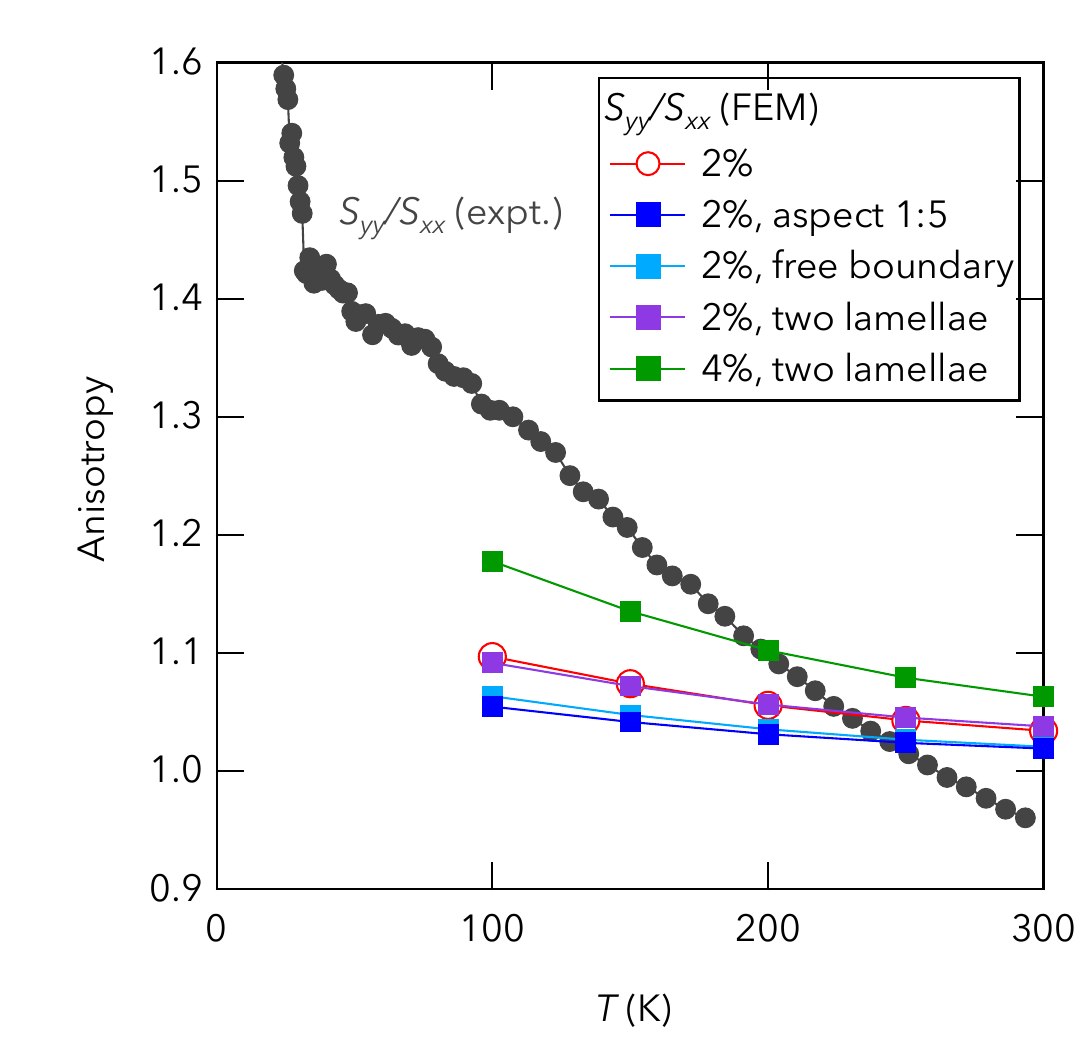}
\caption{
The thermopower anisotropy calculated in various FEM conditions.
The experimental results for the eutectic sample are shown in the black symbol. 
The FEM calculation conditions are listed in Table II.}
\label{figA:ani}
\end{center}
\end{figure}

Based on the calculated strain map of $\varepsilon_{[100]}$ and $\varepsilon_{[010]}$ [Fig.~\ref{fig:fem}(a) and \ref{fig:fem}(b)],
we now calculate the thermopower for the $x$ and $y$ directions in the eutectic system as follows.
We performed DFT and subsequent transport calculations using the deformed unit cells created by varying $\varepsilon_{[100]}$,
and the transport coefficients calculated for several values of $\varepsilon_{[100]}$ are shown in
Figs.~\ref{figA:trans}(a) and \ref{figA:trans}(b).
Note that, owing to computational cost,
the $y$-direction strain $\varepsilon_{[010]}$ is calculated using the linear fit in Fig.~\ref{fig:fem}(c) and  
the $z$-direction strain $\varepsilon_{[001]}$ is obtained using 
the Poisson's ratio
$\varepsilon_{[001]} = -\nu_{xz}\varepsilon_{[100]}$
with $\nu_{xz}=0.21$ \cite{Paglione2002PRB}.
Based on the $\varepsilon_{[100]}$ values for each cell shown in Fig.~\ref{fig:fem}(a), 
the electrical conductivity $\sigma_{ii}$ ($i=x,y$) and thermopower $S_{ii}$ of each cell were determined by interpolating the transport data shown in Figs.~\ref{figA:trans}(a) and \ref{figA:trans}(b).

Using the calculated $\sigma_{ii}$ and $S_{ii}$ for each cell, 
we simulate the total electrical conductivity and the thermopower by employing 
a model combining series and parallel circuits \cite{Lemal2017}.
For the series circuit, the total conductivity $\sigma_{ii}$ ($i=x,y$) and 
the total thermopower $S_{ii}$ are given as 
\begin{align}
\sigma_{ii}^{-1} &= \frac{1}{N}\sum_{n=1}^N\sigma_{ii,n}^{-1},\\
S_{ii} &= \frac{1}{N}\sum_{n=1}^NS_{ii,n},
\end{align}
where $\sigma_{ii,n}$ and $S_{ii,n}$ are 
the electrical conductivity and the thermopower of the $n$-th cell, respectively.
In contrast, for the parallel circuit, these are given as
\begin{align}
\sigma_{ii} &= \frac{1}{N}\sum_{n=1}^N\sigma_{ii,n},\\
\sigma_{ii}S_{ii} &= \frac{1}{N}\sum_{n=1}^N\sigma_{ii,n}S_{ii,n}.
\end{align}
For the Ru inclusion area, 
we used the calculated values of $\sigma_{ii,n}=1.6\times10^{7}$~$\Omega^{-1}$m$^{-1}$ and
$S_{ii,n}/T=-0.11$~nVK$^{-2}$. 
The calculated thermopower is close to the experimental value \cite{Murakami2021}.

\subsection{Sensitivity of the FEM results}

We now examine the sensitivity of the FEM calculations to the condition parameters.
In the main text, we used the following condition: the area fraction of 2\%, aspect ratio of Ru lamella of 1:10 ($x:y$), far-field boundary condition, and one Ru lamella.
We changed these parameters as listed in Table II, 
and the calculated results are shown in Figs.~\ref{figA:femall} and Fig.~\ref{figA:ani}.
The thermopower anisotropy is indeed sensitive to these parameters.
On the other hand, the calculated thermopower anisotropy tends to increase 
with lowering temperature, 
which is qualitatively consistent with the experimental results.

\subsection{Two-component circuit model}

Here we consider a simple two-component circuit model consisting of the \sro crystal and Ru inclusion.
The sample configuration, such as the area fraction and the aspect ratio of the Ru inclusion, is the same as the FEM case.
In this simple model, 
we used the values of the electrical conductivity and the thermopower for the unstrained \sro in Figs.~\ref{figA:trans};
$\sigma_{ii,n}=7.2\times10^{6}$~$\Omega^{-1}$m$^{-1}$ and
$S_{ii,n}/T=0.033$~$\mu$VK$^{-2}$. 
On the other hand, 
the values of the electrical conductivity and the thermopower for the embedded Ru inclusion are 
the same values 
as those used in the FEM calculations:
$\sigma_{ii,n}=1.6\times10^{7}$~$\Omega^{-1}$m$^{-1}$ and $S_{ii,n}/T=-0.11$~nVK$^{-2}$. 
We then calculate the total conductivity and thermopower in the same manner as in the FEM case.
The resulting thermopower anisotropy, which comes from the geometrical arrangement of the rectangular-shaped Ru inclusion,
is almost unity and the deviation is less than 1\%.

\bibliography{ref}

@article{Maeno1994,
    author = {Y. Maeno and H. Hashimoto and K. Yoshida and S. Nishizaki and T. Fujita
              and J. G. Bednorz and F. Lichtenberg},
    title = {Superconductivity in a layered perovskite without copper},
    journal = {Nature},
    volume = {372},
    number = {6506},
    pages = {532--534},
    year = {1994},
    doi = {10.1038/372532a0},
    url = {https://doi.org/10.1038/372532a0},
}

@article{RiceSigrist1995,
    author  = {T. M. Rice and M. Sigrist},
    title   = {{Sr$_2$RuO$_4$: an electronic analogue of $^3$He?}},
    journal = {J. Phys.: Condens. Matter},
    volume  = {7},
    pages   = {L643--L648},
    year    = {1995},
    doi     = {10.1088/0953-8984/7/47/002},
    url     = {https://doi.org/10.1088/0953-8984/7/47/002}
}

@article{Pustogow2019,
    author  = {A. Pustogow and Y. Luo and A. Chronister and Y.-S. Su and D. A. Sokolov and F. Jerzembeck and A. P. Mackenzie and C. W. Hicks and N. Kikugawa and S. Raghu and E. D. Bauer and S. E. Brown},
    title   = {{Constraints on the superconducting order parameter in Sr$_2$RuO$_4$ from oxygen-17 nuclear magnetic resonance}},
    journal = {Nature},
    volume  = {574},
    pages   = {72--75},
    year    = {2019},
    doi     = {10.1038/s41586-019-1596-2},
    url     = {https://doi.org/10.1038/s41586-019-1596-2}
}

@article{Petsch2020,
    author  = {A. N. Petsch and M. Zhu and M. Enderle and Z. Q. Mao and Y. Maeno and I. I. Mazin and S. M. Hayden},
    title   = {{Reduction of the spin susceptibility in the superconducting state of Sr$_2$RuO$_4$ observed by polarized neutron scattering}},
    journal = {Phys. Rev. Lett.},
    volume  = {125},
    pages   = {217004},
    year    = {2020},
    doi     = {10.1103/PhysRevLett.125.217004},
    url     = {https://doi.org/10.1103/PhysRevLett.125.217004}
}

@article{Benhabib2021,
    author  = {S. Benhabib and C. Lupien and I. Paul and L. Berges and M. Dion and M. Nardone and A. Zitouni and Z. Q. Mao and Y. Maeno and A. Georges and L. Taillefer and C. Proust},
    title   = {{Ultrasound evidence for a two-component superconducting order parameter in Sr$_2$RuO$_4$}},
    journal = {Nat. Phys.},
    volume  = {17},
    number  = {2},
    pages   = {194--198},
    year    = {2021},
    doi     = {10.1038/s41567-020-1033-3},
    url     = {https://doi.org/10.1038/s41567-020-1033-3}
}

@article{Grinenko2021,
    author  = {V. Grinenko and S. Ghosh and R. Sarkar and J.-C. Orain and A. Nikitin and M. Elender and D. Das and Z. Guguchia and F. Br{\"u}ckner and M. E. Barber and J. Park and N. Kikugawa and D. A. Sokolov and J. S. Bobowski and T. Miyoshi and Y. Maeno and A. P. Mackenzie and H. Luetkens and C. W. Hicks and H.-H. Klauss},
    title   = {{Split time-reversal symmetry-breaking transitions in Sr$_2$RuO$_4$ under stress}},
    journal = {Nat. Phys.},
    volume  = {17},
    number  = {6},
    pages   = {748--754},
    year    = {2021},
    doi     = {10.1038/s41567-021-01182-7},
    url     = {https://doi.org/10.1038/s41567-021-01182-7}
}

@article{Maeno2024,
    author  = {Y. Maeno and A. Ikeda and G. Mattoni},
    title   = {{Thirty years of puzzling superconductivity in Sr$_2$RuO$_4$}},
    journal = {Nat. Phys.},
    volume  = {20},
    pages   = {1712--1718},
    year    = {2024},
    doi     = {10.1038/s41567-024-02656-0},
    url     = {https://doi.org/10.1038/s41567-024-02656-0}
}

@article{Maeno2024JPSJ,
	author = {Maeno ,Yoshiteru and Yonezawa ,Shingo and Ramires ,Aline},
	date = {2024/06/15},
	doi = {10.7566/JPSJ.93.062001},
	isbn = {0031-9015},
	journal = {J. Phys. Soc. Jpn.},
	month = {2026/05/06},
	number = {6},
	pages = {062001},
	publisher = {The Physical Society of Japan},
	title = {{Still Mystery after All These Years ---Unconventional Superconductivity of Sr$_2$RuO$_4$---}},
	type = {doi: 10.7566/JPSJ.93.062001},
	url = {https://doi.org/10.7566/JPSJ.93.062001},
	volume = {93},
	year = {2024},
	year1 = {2024}}

@article{Maeno1998,
    author = {Y. Maeno and T. Ando and Y. Mori and E. Ohmichi and S. Ikeda
              and S. NishiZaki and S. Nakatsuji},
    title = {{Enhancement of Superconductivity of Sr$_2$RuO$_4$ to 3 K by Embedded Metallic Microdomains}},
    journal = {Phys. Rev. Lett.},
    volume = {81},
    pages = {3765--3768},
    year = {1998},
    doi = {10.1103/PhysRevLett.81.3765},
    url = {https://doi.org/10.1103/PhysRevLett.81.3765},
}

@article{Ando1999,
    author = {T. Ando and T. Akima and Y. Mori and Y. Maeno},
    title = {{Upper Critical Fields of the 3-K Superconducting Phase of Sr$_2$RuO$_4$}},
    journal = {J. Phys. Soc. Jpn.},
    volume = {68},
    number = {6},
    pages = {1651--1654},
    year = {1999},
    doi = {10.1143/JPSJ.68.1651},
    url = {https://doi.org/10.1143/JPSJ.68.1651},
}

@article{Yaguchi2003,
    author = {H. Yaguchi and M. Wada and T. Akima and Y. Maeno and T. Ishiguro},
    title = {{Interface superconductivity in the eutectic Sr$_2$RuO$_4$--Ru: 3-K phase of Sr$_2$RuO$_4$}},
    journal = {Phys. Rev. B},
    volume = {67},
    pages = {214519},
    year = {2003},
    doi = {10.1103/PhysRevB.67.214519},
    url = {https://doi.org/10.1103/PhysRevB.67.214519},
}

@article{Kittaka2009,
    author = {S. Kittaka and T. Nakamura and H. Yaguchi and S. Yonezawa and Y. Maeno},
    title = {{Spatial development of superconductivity in the Sr$_2$RuO$_4$--Ru eutectic system}},
    journal = {J. Phys. Soc. Jpn.},
    volume = {78},
    pages = {064703},
    year = {2009},
    doi = {10.1143/JPSJ.78.064703},
    url = {https://doi.org/10.1143/JPSJ.78.064703},
}

@article{Steppke2017,
    author = {A. Steppke and L. Zhao and M. E. Barber and T. Scaffidi and F. Jerzembeck
              and H. Rosner and A. S. Gibbs and Y. Maeno and S. H. Simon and
              A. P. Mackenzie and C. W. Hicks},
    title = {{Strong peak in $T_{\rm c}$ of Sr$_2$RuO$_4$ under uniaxial pressure}},
    journal = {Science},
    volume = {355},
    pages = {201--204},
    year = {2017},
    doi = {10.1126/science.aaf9398},
    url = {https://doi.org/10.1126/science.aaf9398},
}

@article{Jerzembeck2022,
    author = {F. Jerzembeck and H. S. R{\o}ising and A. Steppke and H. Rosner and 
              D. A. Sokolov and N. Kikugawa and T. Scaffidi and S. H. Simon and 
              A. P. Mackenzie and C. W. Hicks},
    title = {{The superconductivity of Sr$_2$RuO$_4$ under $c$-axis uniaxial stress}},
    journal = {Nat. Commun.},
    volume = {13},
    pages = {4596},
    year = {2022},
    doi = {10.1038/s41467-022-32144-5},
    url = {https://doi.org/10.1038/s41467-022-32144-5},
}

@article{Jerzembeck2023,
    author = {F. Jerzembeck and A. Steppke and A. Pustogow and Y. Luo and 
              A. Chronister and D. A. Sokolov and N. Kikugawa and Y.-S. Li and 
              M. Nicklas and S. E. Brown and A. P. Mackenzie and C. W. Hicks},
    title = {{Upper critical field of Sr$_2$RuO$_4$ under in-plane uniaxial pressure}},
    journal = {Phys. Rev. B},
    volume = {107},
    pages = {064509},
    year = {2023},
    doi = {10.1103/PhysRevB.107.064509},
    url = {https://doi.org/10.1103/PhysRevB.107.064509},
}

@article{Mackenzie2017,
    author = {A. P. Mackenzie and T. Scaffidi and C. W. Hicks and Y. Maeno},
    title = {{Even odder after twenty-three years: the superconducting order parameter puzzle of Sr$_2$RuO$_4$}},
    journal = {npj Quantum Materials},
    volume = {2},
    pages = {40},
    year = {2017},
    doi = {10.1038/s41535-017-0045-4},
    url = {https://doi.org/10.1038/s41535-017-0045-4},
}

@article{Wada2002,
    author  = {M. Wada and H. Yaguchi and M. Yoshioka and Y. Maeno},
    title   = {{Field-temperature phase diagram of Sr$_2$RuO$_4$--Ru eutectic}},
    journal = {J. Phys. Chem. Solids},
    volume  = {63},
    number  = {6--8},
    pages   = {1013--1016},
    year    = {2002},
    doi     = {10.1016/S0022-3697(02)00063-X},
    url     = {https://doi.org/10.1016/S0022-3697(02)00063-X}
}

@article{Sunko2019,
    author = {V. Sunko and E. Abarca Morales and I. Markovic and M. E. Barber 
              and D. Milosavljevic and F. Mazzola and D. A. Sokolov 
              and N. Kikugawa and C. Cacho and P. Dudin and H. Rosner 
              and C. W. Hicks and P. D. C. King and A. P. Mackenzie},
    title = {{Direct observation of a uniaxial stress-driven Lifshitz transition in Sr$_2$RuO$_4$}},
    journal = {npj Quantum Materials},
    volume = {4},
    pages = {46},
    year = {2019},
    doi = {10.1038/s41535-019-0185-9},
    url = {https://doi.org/10.1038/s41535-019-0185-9},
}

@article{Ishida2020,
    author  = {K. Ishida and M. Manago and K. Kinjo and Y. Maeno},
    title   = {{Reduction of the $^{17}$O Knight shift in the superconducting state and the heat-up effect by NMR pulses on Sr$_2$RuO$_4$}},
    journal = {J. Phys. Soc. Jpn.},
    volume  = {89},
    pages   = {034712},
    year    = {2020},
    doi     = {10.7566/JPSJ.89.034712},
    url     = {https://doi.org/10.7566/JPSJ.89.034712}
}

@article{Ghosh2021,
    author  = {S. Ghosh and A. Shekhter and F. Jerzembeck and N. Kikugawa and D. A. Sokolov and M. Brando and A. P. Mackenzie and C. W. Hicks and B. J. Ramshaw},
    title   = {{Thermodynamic evidence for a two-component superconducting order parameter in Sr$_2$RuO$_4$}},
    journal = {Nat. Phys.},
    volume  = {17},
    pages   = {199--204},
    year    = {2021},
    doi     = {10.1038/s41567-020-1032-4},
    url     = {https://doi.org/10.1038/s41567-020-1032-4}
}

@article{Hicks2014,
    author = {C. W. Hicks and D. O. Brodsky and E. Abarca Morales and A. Pustogow 
              and A. S. Gibbs and J. Bruin and M. E. Barber and S. D. Edkins 
              and K. Nishimura and S. Yonezawa and Y. Maeno and A. P. Mackenzie},
    title = {{Strong increase of $T_{\mathrm{c}}$ of Sr$_2$RuO$_4$ under both tensile and compressive strain}},
    journal = {Science},
    volume = {344},
    number = {6181},
    pages = {283--285},
    year = {2014},
    doi = {10.1126/science.1248292},
    url = {https://doi.org/10.1126/science.1248292},
}

@article{Kambara2009,
  author  = {Kambara, H. and Kashiwaya, S. and Yaguchi, H. and Asano, Y. and Tanaka, Y. and Maeno, Y.},
  title   = {{Chiral superconducting phase transition in 3-K phase of Sr$_2$RuO$_4$}},
  journal = {Physica C},
  volume  = {469},
  number  = {15--20},
  pages   = {1030--1033},
  year    = {2009},
  doi     = {10.1016/j.physc.2009.05.176},
  url     = {https://doi.org/10.1016/j.physc.2009.05.176}
}

@article{Paglione2002PRB,
  title = {{Elastic tensor of ${\mathrm{Sr}}_{2}{\mathrm{RuO}}_{4}$}},
  author = {Paglione, Johnpierre and Lupien, C. and MacFarlane, W. A. and Perz, J. M. and Taillefer, Louis and Mao, Z. Q. and Maeno, Y.},
  journal = {Phys. Rev. B},
  volume = {65},
  issue = {22},
  pages = {220506(R)},
  numpages = {4},
  year = {2002},
  month = {May},
  publisher = {American Physical Society},
  doi = {10.1103/PhysRevB.65.220506},
  url = {https://link.aps.org/doi/10.1103/PhysRevB.65.220506}
}

@article{Kouda2022PRR,
    author  = {N. Kouda and K. Eguchi and R. Okazaki and M. Tamura},
    title   = {Anomalous scaling law for thermoelectric transport of two-dimension-confined electrons in an organic molecular system},
    journal = {Phys. Rev. Research},
    volume  = {4},
    pages   = {043050},
    year    = {2022},
    doi     = {10.1103/PhysRevResearch.4.043050},
    url     = {https://doi.org/10.1103/PhysRevResearch.4.043050}
}

@article{Yamanaka2022PRB,
    author  = {T. Yamanaka and R. Okazaki and H. Yaguchi},
    title   = {Enhanced Seebeck coefficient through magnetic fluctuations in {Sr$_2$Ru$_{1-x}$M$_x$O$_4$} ({M} = {Co}, {Mn})},
    journal = {Phys. Rev. B},
    volume  = {105},
    pages   = {184507},
    year    = {2022},
    doi     = {10.1103/PhysRevB.105.184507},
    url     = {https://doi.org/10.1103/PhysRevB.105.184507}
}

@article{Mao2000MRB,
    author  = {Z. Q. Mao and Y. Maeno and H. Fukazawa},
    title   = {Crystal growth of {Sr$_2$RuO$_4$}},
    journal = {Materials Research Bulletin},
    volume  = {35},
    pages   = {1813--1824},
    year    = {2000},
    doi     = {10.1016/S0025-5408(00)00378-0},
    url     = {https://doi.org/10.1016/S0025-5408(00)00378-0}
}

@article{Giannozzi2009JPCM,
    author  = {P. Giannozzi and S. Baroni and N. Bonini and M. Calandra and R. Car and C. Cavazzoni and D. Ceresoli and G. L. Chiarotti and M. Cococcioni and I. Dabo and A. Dal Corso and S. Fabris and G. Fratesi and R. Gebauer and U. Gerstmann and C. Gougoussis and A. Kokalj and M. Lazzeri and L. Martin-Samos and N. Marzari and F. Mauri and R. Mazzarello and S. Paolini and A. Pasquarello and L. Paulatto and C. Sbraccia and S. Scandolo and G. Sclauzero and A. P. Seitsonen and A. Smogunov and P. Umari and R. M. Wentzcovitch},
    title   = {{QUANTUM ESPRESSO}: A modular and open-source software project for quantum simulations of materials},
    journal = {J. Phys.: Condens. Matter},
    volume  = {21},
    pages   = {395502},
    year    = {2009},
    doi     = {10.1088/0953-8984/21/39/395502},
    url     = {https://doi.org/10.1088/0953-8984/21/39/395502}
}

@article{Giannozzi2009JPCM_v2,
    author  = {P. Giannozzi and O. Andreussi and T. Brumme and O. Bunau and M. Buongiorno Nardelli and M. Calandra and R. Car and C. Cavazzoni and D. Ceresoli and M. Cococcioni and N. Colonna and I. Carnimeo and A. Dal Corso and S. de Gironcoli and P. Delugas and R. A. DiStasio Jr. and A. Ferretti and A. Floris and G. Fratesi and G. Fugallo and R. Gebauer and U. Gerstmann and F. Giustino and T. Gorni and J. Jia and M. Kawamura and H.-Y. Ko and A. Kokalj and E. K{\"u}{\c{c}}{\"u}kbenli and M. Lazzeri and M. Marsili and N. Marzari and F. Mauri and N. L. Nguyen and H.-V. Nguyen and A. Otero-de-la-Roza and L. Paulatto and S. Ponc{\'e} and D. Rocca and R. Sabatini and B. Santra and M. Schlipf and A. P. Seitsonen and A. Smogunov and I. Timrov and T. Thonhauser and P. Umari and N. Vast and X. Wu and S. Baroni},
    title   = {Advanced capabilities for materials modelling with {Quantum ESPRESSO}},
    journal = {J. Phys.: Condens. Matter},
    volume  = {29},
    pages   = {465901},
    year    = {2017},
    doi     = {10.1088/1361-648X/aa8f79},
    url     = {https://doi.org/10.1088/1361-648X/aa8f79}
}

@article{Giannozzi2009JPCM_v3,
    author  = {P. Giannozzi and O. Baseggio and P. Bonf{\`a} and D. Brunato and R. Car and I. Carnimeo and C. Cavazzoni and S. de Gironcoli and P. Delugas and F. Ferrari Ruffino and A. Ferretti and N. Marzari and I. Timrov and A. Urru and S. Baroni},
    title   = {{Quantum ESPRESSO} toward the exascale},
    journal = {J. Chem. Phys.},
    volume  = {152},
    pages   = {154105},
    year    = {2020},
    doi     = {10.1063/5.0005082},
    url     = {https://doi.org/10.1063/5.0005082}
}

@article{Kaneyasu2010,
author = {Kaneyasu ,Hirono and Hayashi ,Nobuhiko and Gut ,Bruno and Makoshi ,Kenji and Sigrist ,Manfred},
title = {{Phase Transition in the 3-Kelvin Phase of Eutectic Sr$_2$RuO$_4$-Ru}},
    journal = {J. Phys. Soc. Jpn.},
volume = {79},
number = {10},
pages = {104705},
year = {2010},
doi = {10.1143/JPSJ.79.104705},
URL = { 
        https://doi.org/10.1143/JPSJ.79.104705
},
}

@article{Mao2001,
  title = {{Observation of Andreev Surface Bound States in the 3-K Phase Region of ${\mathrm{Sr}}_{2}{\mathrm{RuO}}_{4}$}},
  author = {Mao, Z. Q. and Nelson, K. D. and Jin, R. and Liu, Y. and Maeno, Y.},
  journal = {Phys. Rev. Lett.},
  volume = {87},
  issue = {3},
  pages = {037003},
  numpages = {4},
  year = {2001},
  month = {Jun},
  publisher = {American Physical Society},
  doi = {10.1103/PhysRevLett.87.037003},
  url = {https://link.aps.org/doi/10.1103/PhysRevLett.87.037003}
}

@article{Kittaka2009L,
	author = {Kittaka ,Shunichiro and Yaguchi ,Hiroshi and Maeno ,Yoshiteru},
	date = {2009/10/15},
	doi = {10.1143/JPSJ.78.103705},
	isbn = {0031-9015},
	journal = {J. Phys. Soc. Jpn.},
	month = {2026/05/11},
	number = {10},
	pages = {103705},
	publisher = {The Physical Society of Japan},
	title = {{Large Enhancement of 3-K Phase Superconductivity in the Sr$_2$RuO$_4$--Ru Eutectic System by Uniaxial Pressure}},
	type = {doi: 10.1143/JPSJ.78.103705},
	url = {https://doi.org/10.1143/JPSJ.78.103705},
	volume = {78},
	year = {2009},
	year1 = {2009}}

@article{Kittaka2010,
  title = {{Higher-${T}_{c}$ superconducting phase in ${\text{Sr}}_{2}{\text{RuO}}_{4}$ induced by uniaxial pressure}},
  author = {Kittaka, Shunichiro and Taniguchi, Haruka and Yonezawa, Shingo and Yaguchi, Hiroshi and Maeno, Yoshiteru},
  journal = {Phys. Rev. B},
  volume = {81},
  issue = {18},
  pages = {180510(R)},
  numpages = {4},
  year = {2010},
  month = {May},
  publisher = {American Physical Society},
  doi = {10.1103/PhysRevB.81.180510},
  url = {https://link.aps.org/doi/10.1103/PhysRevB.81.180510}
}

@article{Taniguchi2014,
	author = {Taniguchi ,Haruka and Nishimura ,Keigo and Goh ,Swee K. and Yonezawa ,Shingo and Maeno ,Yoshiteru},
	date = {2015/01/15},
	doi = {10.7566/JPSJ.84.014707},
	isbn = {0031-9015},
	journal = {J. Phys. Soc. Jpn.},
	month = {2026/05/11},
	number = {1},
	pages = {014707},
	publisher = {The Physical Society of Japan},
	title = {{Higher-$T_{\rm c}$ Superconducting Phase in Sr$_2$RuO$_4$ Induced by In-Plane Uniaxial Pressure}},
	type = {doi: 10.7566/JPSJ.84.014707},
	url = {https://doi.org/10.7566/JPSJ.84.014707},
	volume = {84},
	year = {2014},
	year1 = {2014}}

@article{Ying2009,
  title = {{Suppression of Proximity Effect and the Enhancement of $p$-Wave Superconductivity in the ${\mathrm{Sr}}_{2}{\mathrm{RuO}}_{4}$-Ru System}},
  author = {Ying, Y. A. and Xin, Y. and Clouser, B. W. and Hao, E. and Staley, N. E. and Myers, R. J. and Allard, L. F. and Fobes, D. and Liu, T. and Mao, Z. Q. and Liu, Y.},
  journal = {Phys. Rev. Lett.},
  volume = {103},
  issue = {24},
  pages = {247004},
  numpages = {4},
  year = {2009},
  month = {Dec},
  publisher = {American Physical Society},
  doi = {10.1103/PhysRevLett.103.247004},
  url = {https://link.aps.org/doi/10.1103/PhysRevLett.103.247004}
}

@article{Mackenzie1998,
  title = {{Extremely Strong Dependence of Superconductivity on Disorder in ${\mathrm{Sr}}_{2}{\mathrm{RuO}}_{4}$}},
  author = {Mackenzie, A. P. and Haselwimmer, R. K. W. and Tyler, A. W. and Lonzarich, G. G. and Mori, Y. and Nishizaki, S. and Maeno, Y.},
  journal = {Phys. Rev. Lett.},
  volume = {80},
  issue = {1},
  pages = {161--164},
  numpages = {0},
  year = {1998},
  month = {Jan},
  publisher = {American Physical Society},
  doi = {10.1103/PhysRevLett.80.161},
  url = {https://link.aps.org/doi/10.1103/PhysRevLett.80.161}
}

@article{Ohsumi2024,
  title = {{Transverse Thermoelectric Conversion in the Mixed-Dimensional Semimetal $\mathrm{WSi}$${}_{2}$}},
  author = {Ohsumi, Shoya and Sato, Yoshiki J. and Okazaki, Ryuji},
  journal = {PRX Energy},
  volume = {3},
  issue = {4},
  pages = {043007},
  numpages = {12},
  year = {2024},
  month = {Nov},
  publisher = {American Physical Society},
  doi = {10.1103/PRXEnergy.3.043007},
  url = {https://link.aps.org/doi/10.1103/PRXEnergy.3.043007}
}

@article{Nishinakayama2024,
  title = {{Negatively enhanced thermopower near a Van Hove singularity in electron-doped ${\mathrm{Sr}}_{2}{\mathrm{RuO}}_{4}$}},
  author = {Nishinakayama, Rei and Sato, Yoshiki J. and Yamanaka, Takayoshi and Maeno, Yoshiteru and Yaguchi, Hiroshi and Kikugawa, Naoki and Okazaki, Ryuji},
  journal = {Phys. Rev. Res.},
  volume = {6},
  issue = {2},
  pages = {023222},
  numpages = {10},
  year = {2024},
  month = {May},
  publisher = {American Physical Society},
  doi = {10.1103/PhysRevResearch.6.023222},
  url = {https://link.aps.org/doi/10.1103/PhysRevResearch.6.023222}
}

@article{Yoshino1996,
	author = {Yoshino ,Harukazu and Murata ,Keizo and Shirakawa ,Naoki and Nishihara ,Yoshikazu and Maeno ,Yoshiteru and Fujita ,Toshizo},
	date = {1996/06/15},
	doi = {10.1143/JPSJ.65.1548},
	isbn = {0031-9015},
	journal = {J. Phys. Soc. Jpn.},
	month = {2026/05/13},
	number = {6},
	pages = {1548--1550},
	publisher = {The Physical Society of Japan},
	title = {{Thermopower of a Layered Perovskite Superconductor, Sr$_2$RuO$_4$}},
	type = {doi: 10.1143/JPSJ.65.1548},
	url = {https://doi.org/10.1143/JPSJ.65.1548},
	volume = {65},
	year = {1996},
	year1 = {1996}}

@article{Xu2008,
  title = {{Band-Dependent Normal-State Coherence in ${\mathrm{Sr}}_{2}{\mathrm{RuO}}_{4}$: Evidence from Nernst Effect and Thermopower Measurements}},
  author = {Xu, X. F. and Xu, Z. A. and Liu, T. J. and Fobes, D. and Mao, Z. Q. and Luo, J. L. and Liu, Y.},
  journal = {Phys. Rev. Lett.},
  volume = {101},
  issue = {5},
  pages = {057002},
  numpages = {4},
  year = {2008},
  month = {Jul},
  publisher = {American Physical Society},
  doi = {10.1103/PhysRevLett.101.057002},
  url = {https://link.aps.org/doi/10.1103/PhysRevLett.101.057002}
}

@article{Daou2023,
  title = {{Anisotropic Seebeck coefficient of ${\mathrm{Sr}}_{2}{\mathrm{RuO}}_{4}$ in the incoherent regime}},
  author = {Daou, Ramzy and H\'ebert, Sylvie and Grissonnanche, Ga\"el and Hassinger, Elena and Taillefer, Louis and Taniguchi, Haruka and Maeno, Yoshiteru and Gibbs, Alexandra S. and Mackenzie, Andrew P.},
  journal = {Phys. Rev. B},
  volume = {108},
  issue = {12},
  pages = {L121106},
  numpages = {5},
  year = {2023},
  month = {Sep},
  publisher = {American Physical Society},
  doi = {10.1103/PhysRevB.108.L121106},
  url = {https://link.aps.org/doi/10.1103/PhysRevB.108.L121106}
}

@article{Otsuki2023,
  title = {Carrier filtering effect for enhanced thermopower in a body-centered tetragonal ruthenate},
  author = {Otsuki, Ryota and Sato, Yoshiki J. and Okazaki, Ryuji and Komine, Tomoya and Kurihara, Ryosuke and Yaguchi, Hiroshi},
  journal = {Phys. Rev. Mater.},
  volume = {7},
  issue = {12},
  pages = {125401},
  numpages = {8},
  year = {2023},
  month = {Dec},
  publisher = {American Physical Society},
  doi = {10.1103/PhysRevMaterials.7.125401},
  url = {https://link.aps.org/doi/10.1103/PhysRevMaterials.7.125401}
}

@article{Kawamura2014,
  title = {{Improved tetrahedron method for the Brillouin-zone integration applicable to response functions}},
  author = {Kawamura, Mitsuaki and Gohda, Yoshihiro and Tsuneyuki, Shinji},
  journal = {Phys. Rev. B},
  volume = {89},
  issue = {9},
  pages = {094515},
  numpages = {8},
  year = {2014},
  month = {Mar},
  publisher = {American Physical Society},
  doi = {10.1103/PhysRevB.89.094515},
  url = {https://link.aps.org/doi/10.1103/PhysRevB.89.094515}
}

@article{Kawamura2019,
	author = {Kawamura, Mitsuaki},
	date = {2019/06/01/},
	doi = {https://doi.org/10.1016/j.cpc.2019.01.017},
	isbn = {0010-4655},
	journal = {Comput. Phys. Commun.},
	pages = {197--203},
	title = {{FermiSurfer: Fermi-surface viewer providing multiple representation schemes}},
	url = {https://www.sciencedirect.com/science/article/pii/S0010465519300347},
	volume = {239},
	year = {2019}}

@article{Chmaissem1998,
  title = {{Thermal expansion and compressibility of ${\mathrm{Sr}}_{2}{\mathrm{RuO}}_{4}$}},
  author = {Chmaissem, O. and Jorgensen, J. D. and Shaked, H. and Ikeda, S. and Maeno, Y.},
  journal = {Phys. Rev. B},
  volume = {57},
  issue = {9},
  pages = {5067--5070},
  numpages = {0},
  year = {1998},
  month = {Mar},
  publisher = {American Physical Society},
  doi = {10.1103/PhysRevB.57.5067},
  url = {https://link.aps.org/doi/10.1103/PhysRevB.57.5067}
}

@article{Mravlje2016,
  title = {{Thermopower and Entropy: Lessons from ${\mathrm{Sr}}_{2}{\mathrm{RuO}}_{4}$}},
  author = {Mravlje, Jernej and Georges, Antoine},
  journal = {Phys. Rev. Lett.},
  volume = {117},
  issue = {3},
  pages = {036401},
  numpages = {5},
  year = {2016},
  month = {Jul},
  publisher = {American Physical Society},
  doi = {10.1103/PhysRevLett.117.036401},
  url = {https://link.aps.org/doi/10.1103/PhysRevLett.117.036401}
}

@article{Madsen2006,
	author = {Madsen, Georg K. H. and Singh, David J.},
	date = {2006/07/01/},
	doi = {https://doi.org/10.1016/j.cpc.2006.03.007},
	isbn = {0010-4655},
	journal = {Comput. Phys. Commun.},
	number = {1},
	pages = {67--71},
	title = {{BoltzTraP. A code for calculating band-structure dependent quantities}},
	url = {https://www.sciencedirect.com/science/article/pii/S0010465506001305},
	volume = {175},
	year = {2006}}

@article{Finkel1971,
  title = {{X-ray diffraction study of the thermal expansion of Ru, Os and Re at 77-300 K}},
  author = {V. A. Finkel and M. I. Palatnik and G. P. Kovtun},
  journal = {Phys. Met. Metallogr. (USSR)},
  volume = {32},
  pages = {231},
  year = {1971},
}

@article{Manako2024,
	author = {Manako, Hikari and Ohsumi, Shoya and Sato, Yoshiki J. and Okazaki, R. and Aoki, D.},
	date = {2024/05/09},
	doi = {10.1038/s41467-024-48217-0},
	id = {Manako2024},
	isbn = {2041-1723},
	journal = {Nat. Commun.},
	number = {1},
	pages = {3907},
	title = {{Large transverse thermoelectric effect induced by the mixed-dimensionality of Fermi surfaces}},
	url = {https://doi.org/10.1038/s41467-024-48217-0},
	volume = {15},
	year = {2024}}

@article{Herman2019,
  title = {{Deviation from Fermi-liquid transport behavior in the vicinity of a Van Hove singularity}},
  author = {Herman, Franti\ifmmode \check{s}\else \v{s}\fi{}ek and Buhmann, Jonathan and Fischer, Mark H. and Sigrist, Manfred},
  journal = {Phys. Rev. B},
  volume = {99},
  issue = {18},
  pages = {184107},
  numpages = {10},
  year = {2019},
  month = {May},
  publisher = {American Physical Society},
  doi = {10.1103/PhysRevB.99.184107},
  url = {https://link.aps.org/doi/10.1103/PhysRevB.99.184107}
}

@article{Buhmann2013,
  title = {{Thermoelectric effect of correlated metals: Band-structure effects and the breakdown of Mott's formula}},
  author = {Buhmann, Jonathan M. and Sigrist, Manfred},
  journal = {Phys. Rev. B},
  volume = {88},
  issue = {11},
  pages = {115128},
  numpages = {8},
  year = {2013},
  month = {Sep},
  publisher = {American Physical Society},
  doi = {10.1103/PhysRevB.88.115128},
  url = {https://link.aps.org/doi/10.1103/PhysRevB.88.115128}
}

@article{Pickett1988,
  title = {{Smooth Fourier interpolation of periodic functions}},
  author = {Pickett, Warren E. and Krakauer, Henry and Allen, Philip B.},
  journal = {Phys. Rev. B},
  volume = {38},
  issue = {4},
  pages = {2721--2726},
  numpages = {0},
  year = {1988},
  month = {Aug},
  publisher = {American Physical Society},
  doi = {10.1103/PhysRevB.38.2721},
  url = {https://link.aps.org/doi/10.1103/PhysRevB.38.2721}
}

@article{Barber2018,
  title = {{Resistivity in the Vicinity of a van Hove Singularity: ${\mathrm{Sr}}_{2}{\mathrm{RuO}}_{4}$ under Uniaxial Pressure}},
  author = {Barber, M. E. and Gibbs, A. S. and Maeno, Y. and Mackenzie, A. P. and Hicks, C. W.},
  journal = {Phys. Rev. Lett.},
  volume = {120},
  issue = {7},
  pages = {076602},
  numpages = {6},
  year = {2018},
  month = {Feb},
  publisher = {American Physical Society},
  doi = {10.1103/PhysRevLett.120.076602},
  url = {https://link.aps.org/doi/10.1103/PhysRevLett.120.076602}
}

@article{Georges2021,
  title = {{Skewed non-Fermi liquids and the Seebeck effect}},
  author = {Georges, Antoine and Mravlje, Jernej},
  journal = {Phys. Rev. Res.},
  volume = {3},
  issue = {4},
  pages = {043132},
  numpages = {12},
  year = {2021},
  month = {Nov},
  publisher = {American Physical Society},
  doi = {10.1103/PhysRevResearch.3.043132},
  url = {https://link.aps.org/doi/10.1103/PhysRevResearch.3.043132}
}

@article{Aramberri2017,
  title = {{Strain effects in topological insulators: Topological order and the emergence of switchable topological interface states in ${\mathbf{Sb}}_{2}{\mathbf{Te}}_{3}/{\mathbf{Bi}}_{2}{\mathbf{Te}}_{3}$ heterojunctions}},
  author = {Aramberri, H. and Mu\~noz, M. C.},
  journal = {Phys. Rev. B},
  volume = {95},
  issue = {20},
  pages = {205422},
  numpages = {11},
  year = {2017},
  month = {May},
  publisher = {American Physical Society},
  doi = {10.1103/PhysRevB.95.205422},
  url = {https://link.aps.org/doi/10.1103/PhysRevB.95.205422}
}

@article{Kawasugi2019,
	author = {Kawasugi, Yoshitaka and Seki, Kazuhiro and Tajima, Satoshi and Pu, Jiang and Takenobu, Taishi and Yunoki, Seiji and Yamamoto, Hiroshi M. and Kato, Reizo},
	doi = {10.1126/sciadv.aav7282},
	journal = {Sci. Adv.},
	month = {2026/05/25},
	number = {5},
    year = {2019},
	pages = {eaav7282},
	publisher = {American Association for the Advancement of Science},
	title = {{Two-dimensional ground-state mapping of a Mott-Hubbard system in a flexible field-effect device}},
	type = {doi: 10.1126/sciadv.aav7282},
	url = {https://doi.org/10.1126/sciadv.aav7282},
	volume = {5}}

@article{Sootsman2010,
	author = {Sootsman, Joseph R. and He, Jiaqing and Dravid, Vinayak P. and Ballikaya, Sedat and Vermeulen, Derek and Uher, Ctirad and Kanatzidis, Mercouri G.},
	date = {2010/02/09},
	doi = {10.1021/cm9016672},
	isbn = {0897-4756},
	journal = {Chem. Mater.},
	month = {02},
	number = {3},
	pages = {869--875},
	publisher = {American Chemical Society},
	title = {{Microstructure and Thermoelectric Properties of Mechanically Robust PbTe-Si Eutectic Composites}},
	type = {doi: 10.1021/cm9016672},
	url = {https://doi.org/10.1021/cm9016672},
	volume = {22},
	year = {2010},
	year1 = {2010}}

@article{Lemal2017,
  title = {{Thermoelectric properties of layered calcium cobaltite ${\mathrm{Ca}}_{3}{\mathrm{Co}}_{4}{\mathrm{O}}_{9}$ from hybrid functional first-principles calculations}},
  author = {Lemal, S\'ebastien and Varignon, Julien and Bilc, Daniel I. and Ghosez, Philippe},
  journal = {Phys. Rev. B},
  volume = {95},
  issue = {7},
  pages = {075205},
  numpages = {14},
  year = {2017},
  month = {Feb},
  publisher = {American Physical Society},
  doi = {10.1103/PhysRevB.95.075205},
  url = {https://link.aps.org/doi/10.1103/PhysRevB.95.075205}
}

@article{Murakami2021,
title = {{Mechanical and thermoelectric properties of iridium-ruthenium alloy grown by the micro-pulling-down method}},
journal = {J. Cryst. Growth},
volume = {573},
pages = {126256},
year = {2021},
issn = {0022-0248},
doi = {https://doi.org/10.1016/j.jcrysgro.2021.126256},
url = {https://www.sciencedirect.com/science/article/pii/S0022024821002311},
author = {Rikito Murakami and Kei Kamada and Katsunari Oikawa and Akira Yoshikawa},
}

@article{Damascelli2000,
  title = {{Fermi Surface, Surface States, and Surface Reconstruction in ${\mathrm{Sr}}_{2}{\mathrm{RuO}}_{4}$}},
  author = {Damascelli, A. and Lu, D. H. and Shen, K. M. and Armitage, N. P. and Ronning, F. and Feng, D. L. and Kim, C. and Shen, Z.-X. and Kimura, T. and Tokura, Y. and Mao, Z. Q. and Maeno, Y.},
  journal = {Phys. Rev. Lett.},
  volume = {85},
  issue = {24},
  pages = {5194--5197},
  numpages = {0},
  year = {2000},
  month = {Dec},
  publisher = {American Physical Society},
  doi = {10.1103/PhysRevLett.85.5194},
  url = {https://link.aps.org/doi/10.1103/PhysRevLett.85.5194}
}

@article{Bergemann2000,
  title = {{Detailed Topography of the Fermi Surface of ${\mathrm{Sr}}_{2}{\mathrm{RuO}}_{4}$}},
  author = {Bergemann, C. and Julian, S. R. and Mackenzie, A. P. and NishiZaki, S. and Maeno, Y.},
  journal = {Phys. Rev. Lett.},
  volume = {84},
  issue = {12},
  pages = {2662--2665},
  numpages = {0},
  year = {2000},
  month = {Mar},
  publisher = {American Physical Society},
  doi = {10.1103/PhysRevLett.84.2662},
  url = {https://link.aps.org/doi/10.1103/PhysRevLett.84.2662}
}

@article{Mackenzie1996,
  title = {{Quantum Oscillations in the Layered Perovskite Superconductor S${\mathrm{r}}_{2}$Ru${\mathrm{O}}_{4}$}},
  author = {Mackenzie, A. P. and Julian, S. R. and Diver, A. J. and McMullan, G. J. and Ray, M. P. and Lonzarich, G. G. and Maeno, Y. and Nishizaki, S. and Fujita, T.},
  journal = {Phys. Rev. Lett.},
  volume = {76},
  issue = {20},
  pages = {3786--3789},
  numpages = {0},
  year = {1996},
  month = {May},
  publisher = {American Physical Society},
  doi = {10.1103/PhysRevLett.76.3786},
  url = {https://link.aps.org/doi/10.1103/PhysRevLett.76.3786}
}

@article{Deeks2002,
	author = {Deeks, A. J. and Wolf, J. P.},
	date = {2002/06/01},
	doi = {10.1007/s00466-002-0314-2},
	id = {Deeks2002},
	isbn = {1432-0924},
	journal = {Comput. Mechanics},
	number = {6},
	pages = {489--504},
	title = {{A virtual work derivation of the scaled boundary finite-element method for elastostatics}},
	url = {https://doi.org/10.1007/s00466-002-0314-2},
	volume = {28},
	year = {2002}}

@article{Momma2011,
    author = {Momma, Koichi and Izumi, Fujio},
    url = {https://doi.org/10.1107/S0021889811038970},
    doi = {10.1107/S0021889811038970},
    journal = {J. Appl. Crystal.},
    month = {Dec},
    number = {6},
    pages = {1272--1276},
    title = {{{\it VESTA3} for three-dimensional visualization of crystal, volumetric and morphology data}},
    volume = {44},
    year = {2011},
}

@article{Beekman2024,
    author = {Beekman, Matt and Benedetti, Matteo and Dominguez, Deja and Hewett-Abbott, Hayden and Jarymowycz, Andrew and Leibowitz, Matthew and Nichols, Travis and Dorris, Roger and Thomson, Kyle and Watzman, Sarah J. and Gibson, Thu and Schlaak, Katherine A.},
    title = {{Validation of galvanomagnetic and thermomagnetic transport measurements using Standard Reference Material 3451}},
    journal = {Rev. Sci. Instrum.},
    volume = {95},
    number = {11},
    pages = {113905},
    year = {2024},
    month = {11},
    issn = {0034-6748},
    doi = {10.1063/5.0235568},
    url = {https://doi.org/10.1063/5.0235568},
}

@book{Shklovskii,
    author = {B. I. Shklovskii and A. L.  Efros},
    title = {{Electronic Properties of Doped Semiconductors}},
    year = {1984},
    publisher = {Springer, New York},
}

\end{document}